\documentclass[preprint,authoryear,review,12pt]{elsarticle}

\usepackage[version=3]{mhchem}
\usepackage{amssymb}
\usepackage{longtable}
\usepackage{setspace}
\usepackage{xspace} 
\usepackage{wasysym}
\usepackage{lscape}
\usepackage{fancyhdr}
\usepackage{fullpage}
\usepackage{caption}
\usepackage{graphicx, subfigure}
\usepackage{array}
\usepackage{rotating}
\usepackage{dcolumn}
\usepackage{natbib}
\usepackage{booktabs}
\usepackage[greek,english]{babel}

\usepackage{outlines}

\newcommand\TT{\rule{0pt}{3.5ex}}
\newcommand\BB{\rule[-2.2ex]{0pt}{0pt}}

\newcommand{\delox}{\greektext d\latintext \ce{^{18}O}\xspace}
\newcommand{\deloy}{\greektext d\latintext \ce{^{17}O}\xspace}
\newcommand{\deloxy}{\greektext d\latintext \ce{^{17,18}O}\xspace}
\newcommand{\Deloy}{\greektext D\latintext \ce{^{17}O}\xspace}

\newcolumntype{L}{D{.}{.}{2,5}}

\journal{Geochimica et Cosmochimica Acta}

\begin{document}

\begin{frontmatter}



\title{Oxygen Isotopic Composition of Coarse- and Fine-grained Material from Comet 81P/Wild 2}


\author[l1]{Ryan C. Ogliore}
\author[l1]{Kazuhide Nagashima}
\author[l1]{Gary R. Huss}

\address[l1]{Hawai`i Institute of Geophysics and Planetology, University of Hawai`i at M\={a}noa, Honolulu, HI 96822, USA}

\author[l2]{Andrew J. Westphal}
\author[l2]{Zack Gainsforth}
\author[l2]{Anna L. Butterworth}

\address[l2]{Space Sciences Laboratory, University of California at Berkeley, Berkeley, CA 94720, USA}

\begin{keyword}
comets \sep solar nebula \sep Jupiter-family comets \sep Wild 2 \sep oxygen isotopes \sep cosmochemistry
\end{keyword}

\begin{abstract}

Individual particles from comet 81P/Wild 2 collected by NASA's Stardust mission vary in size from small sub-$\mu$m fragments found in the walls of the aerogel tracks, to large fragments up to tens of $\mu$m in size found towards the termini of tracks. The comet, in an orbit beyond Neptune since its formation, retains an intact a record of early-Solar-System processes that was compromised in asteroidal samples by heating and aqueous alteration. We measured the O isotopic composition of seven Stardust fragments larger than $\sim$2 $\mu$m extracted from five different Stardust aerogel tracks, and 63 particles smaller than $\sim$2 $\mu$m from the wall of a Stardust track. The larger particles show a relatively narrow range of O isotopic compositions that is consistent with \ce{^{16}O}-poor phases commonly seen in meteorites.  Many of the larger Stardust fragments studied so far have chondrule-like mineralogy which is consistent with formation in the inner Solar System.  The fine-grained material shows a very broad range of O isotopic compositions ($-70$\permil\xspace$<$  \Deloy$<+60$\permil) suggesting that Wild 2 fines are either primitive outer-nebula dust or a very diverse sampling of inner Solar System compositional reservoirs that accreted along with a large number of inner-Solar-System rocks to form comet Wild 2.

\end{abstract}

\end{frontmatter}

\section{Introduction}

Oxygen isotopic compositions are traditionally reported in delta notation --- the parts-per-thousand deviations of the stable O isotope ratios from standard mean ocean water (($^{18}$O/$^{16}$O)$_{\textnormal{SMOW}}$ =  0.0020052 and ($^{17}$O/$^{16}$O)$_{\textnormal{SMOW}}$ =  0.00038288):
%\citep{baertschi1976absolute}:

\begin{eqnarray}
\textnormal{\delox} =& \left(\frac{^{18}\textnormal{O}/^{16}\textnormal{O}}{(^{18}\textnormal{O}/^{16}\textnormal{O})_{\textnormal{SMOW}}} - 1\right)\times1000\\
\textnormal{\deloy} =& \left(\frac{^{17}\textnormal{O}/^{16}\textnormal{O}}{(^{17}\textnormal{O}/^{16}\textnormal{O})_{\textnormal{SMOW}}} - 1\right)\times1000
\end{eqnarray}

% We also use the standard definition of \Deloy, the difference between the measured \deloy and the (mass-dependent) terrestrial fractionation line:
% \begin{equation}
% \textnormal{\Deloy} = \textnormal{\deloy} - 0.52\times\textnormal{\delox}
% \end{equation}

Most terrestrial processes that fractionate one isotope of an element relative to another are due to the fact that vibrational frequencies, bond strengths, and diffusion coefficients depend on an isotope's mass. Such mass-dependent fractionation, arising from equilibrium or kinetic effects, causes $^{18}$O/$^{16}$O to change approximately twice as much as $^{17}$O/$^{16}$O. On an oxygen three-isotope plot of \deloy vs. \delox, most terrestrial samples fall along a line of slope $\sim$0.52, called the terrestrial fractionation line (TFL). However, some extraterrestrial samples do not follow this trend. Calcium-aluminum inclusions, chondrules, and amoeboid olivine aggregates found in primitive chondritic meteorites fall on a slope $\sim$1 line called the carbonaceous chondrite anhydrous mineral (CCAM) line \citep{clayton1977distribution}. The difference between the measured \deloy and the TFL is defined as:

\begin{equation}
\textnormal{\Deloy} = \textnormal{\deloy} - 0.52\times\textnormal{\delox}
\end{equation}

Refractory components like calcium-aluminum inclusions typically have a lower (i.e. more negative) \Deloy than chondrules (low \Deloy objects are also called ``$^{16}$O-rich''). However, most rocky inner Solar System materials (Earth, Moon, Mars, and bulk asteroids) are close to \deloxy=(0 , 0). The O composition of the Sun, inferred from measurements of returned samples of the solar wind, is very $^{16}$O-rich: \deloxy=($−59.1$\permil , $−58.5$\permil) \citep{mckeegan2011oxygen}. Photochemical processing \citep[e.g.][]{clayton2002solar} has been suggested to  explain the CCAM line by mass-independent isotope fractionation (processes that enrich $^{18}$O/$^{16}$O and $^{17}$O/$^{16}$O equally), but the large difference between the $^{16}$O-rich Sun and the $^{16}$O-poor inner Solar System solids would require such a process to be extremely efficient. It is more plausible that different reservoirs of O isotopic compositions were inherited from the Sun's parent molecular cloud \citep{huss2012differences}, and the CCAM line reflects mixing between reservoirs. Oxygen isotopic measurements of primitive Solar System materials could yield insight into the composition of the dust and gas making up the protosolar molecular cloud.

The Moon, terrestrial planets, and large asteroids underwent melting and differentiation due to accretional and radiogenic heating, so the O isotopic signature of the Solar System's building blocks are lost in these objects. Some ``primitive'' meteorites from smaller asteroids avoided significant aqueous and thermal processing. Analyses of these samples give us insight into, for example, the high-temperature refractory objects that were the first to form in the solar nebula, and the inner Solar System abundance of isotopically anomalous dust which formed around other stars (presolar grains).

However, outer Solar System bodies may retain pristine dust from the solar nebula along with ices and other volatile species that were vaporized closer to the Sun. Jupiter-family comets, in cold storage nearly all of their lives, preserved the O isotopic composition of the materials that originally accreted to form the comet in the outer Solar System. The diversity of O isotopic reservoirs inherited from the protosolar molecular cloud may still be found in cometary particles. Additionally, cometary material may contain a higher abundance, or additional types, of presolar grains than what is seen in primitive meteorites. More fragile presolar grains can be easily destroyed in the solar nebula or in an asteroid; the abundance and types of presolar grains in comets may reflect the true contribution of this exotic material to the solar nebula.

NASA's Stardust mission returned samples from the Jupiter-family comet 81P/Wild 2 in 2006 for laboratory study \citep{brownlee2006comet}. Impacting cometary particles were collected at $\sim$6 km/s in aerogel, which acted somewhat as a size-sorting sieve, with larger and more robust rock fragments, called terminal particles, found at the ends of tracks, while fine-grained material is found closer to the space-exposed aerogel surface in the track bulb. The terminal particles, protected by their thermal inertia, were collected mostly intact whereas the fines were intimately mixed with aerogel \citep[e.g.][]{Stodolna:2012p5651}.

The terminal particles are the easiest objects to study due to their larger size, better preservation, and easier sample preparation. Many of the terminal particles studied so far have a high-temperature origin, likely in the inner Solar System: they are large ($>$5 $\mu$m) single crystals \citep[e.g. Track C2115,22,20: terminal particle,][]{Joswiak:2012p6266}, or have mineralogy consistent with fragments of chondrules \citep[e.g. Track C2054,0,35,6: ``Torajiro'',][]{Nakamura:2008p2349}, calcium-aluminum inclusions \citep[e.g. Track C2054,4,25: ``Inti'',][]{Simon:2008p2065}, metal \citep[e.g. Track C2044,32,41: ``Simieo'',][]{2012M&PSA..75.5309W}, and other phases. 

Though a relatively small number of analyses have been performed on comet Wild 2 samples, it appears that the large terminal particles show more compositional diversity than similar objects in a given type of meteorite. Measurements of the FeO and MnO composition of Wild 2 ferrous olivine do not show a clear linear relationship as do ferrous olivines from specific classes of meteorites \citep{Brownlee:2012p6345}, indicating that Wild 2 olivines could not have originated from a small number of chondrite parent bodies. While the presolar grain abundance of comet Wild 2 has been difficult to precisely determine due to the small sample size and capture effects, the large majority of Wild 2 particles do not show extreme isotopic anomalies. The oxygen isotopic composition of $>$3 $\mu$m ferromagnesian crystalline silicates in Wild 2 show a \Deloy dependence on Mg content that is similar to CR chondrite chondrules \citep{nakashima2012oxygen}, though Wild 2 has a much larger relative abundance of FeO-rich particles than CR chondrites \citep{Zolensky:2006p1378}. 

Asteroids accreted local material of relatively homogenous composition, as demonstrated by the diagnostic properties of meteorites (e.g. chondrule size, matrix fraction, oxygen isotopic composition) that allows different meteorites to be grouped together as originating from the same parent asteroid. Conversely, the larger particles in comet Wild 2 appear to have formed far from where the
parent body accreted \citep{Brownlee:2012p5596}, that is, material was transported from the inner Solar System \citep[e.g.][]{ciesla2007outward} to beyond the orbit of Neptune where Jupiter-family comets like Wild 2 probably formed.

%the scattered disk at $\sim$35 AU \citep[e.g.][]{ciesla2007outward} where the comet likely formed \citep{Duncan:1997p5478}.

The chemical compositions of the fine-grained material located in the bulbous cavity of a Stardust track have been analyzed by \citet{Stodolna:2012p5651}. Despite difficulties arising from sample alteration due to high-speed capture into aerogel, the fine-grained bulb material appears to be approximately solar in its average Fe-Mg-S composition, but heterogeneous on a particle-by-particle basis \citep{Westphal:2009p1874}. These are the expected characteristics of primitive dust that has not been significantly thermally altered in the solar nebula or in a parent body.

Fine-grained cometary dust captured in the track bulb could be the missing ``pristine'' nebular material that was expected to make up the bulk of comet Wild 2 before the Stardust mission returned with samples. The fines are either (1) outer-nebula dust, locally present in the outer Solar System where comet Wild 2 formed and likely representative of unprocessed particles that made up the Solar System's parent molecular cloud, or (2) inner Solar System material, possibly related to meteorite matrix, that was transported to the outer disk along with the terminal particles, or a combination of (1) and (2). Fines from source 1) would be rich in circumstellar grains with nucleosynthetic isotopic anomalies and other material inherited from the Solar System's parent molecular cloud, whereas those from source (2) would be similar in composition to the fine-grained matrix of primitive meteorites (material that did not experience thermal or aqueous processing on its asteroidal parent body). 

To gain insight into the origins of coarse-grained ($>$2 $\mu$m) and fine-grained ($<$2 $\mu$m) material from comet Wild 2, we measured the oxygen isotopic composition of terminal particles and fine-grained bulb material of cometary matter returned by NASA's Stardust mission. 

\section{Description of Wild 2 Particles}
\subsection{Individual fragments}
\label{coarsedescrip}
We characterized the mineralogy and measured the O isotopic composition of seven Stardust fragments larger than $\sim$2 $\mu$m extracted from five different Stardust aerogel tracks. The seven fragments are described individually below and the tracks from which they were extracted are described in Table \ref{TrackTable}. The pyroxene and olivine phases we analyzed all had relatively FeO-rich compositions, i.e. all had mol. \% fayalite $>$10. This lack of MgO-rich silicates is not surprising, since it has been shown \citep[e.g., ][]{frank2014olivine} that Wild 2 has a flatter distribution of fayalite content compared to matrix grains in chondrites, and lacks the MgO-rich silicates (mol. \% fayalite $<$5) that are abundant in many meteorite classes.

\setlength{\tabcolsep}{6pt}
\begin{table}[!ht]
\centering
\begin{tabular}{lccc}
\hline  \TT \BB
Track&Type$^{\dag}$&Length (mm)&Particles Analyzed\\
\hline\\[0.5ex]
C2052,74 & B & 8.0 & Iris, Callie, Tintin, bulb material\\
C2061,113 & B & 1.5 & Pooka\\
C2035,105 & B & 2.2& Caligula\\
C2009,77 & B & 2.5& The Butterfly\\
C2062,162 & A &2.9 & Cecil\\[1ex]
\hline
\end{tabular}
\caption{Track Information. $^{\dag}$Stardust cometary track types are described by \citet{burchell2008characteristics}. \label{TrackTable}}
\end{table}

\begin{figure}[!ht]
\begin{center}
\includegraphics[width=\textwidth]{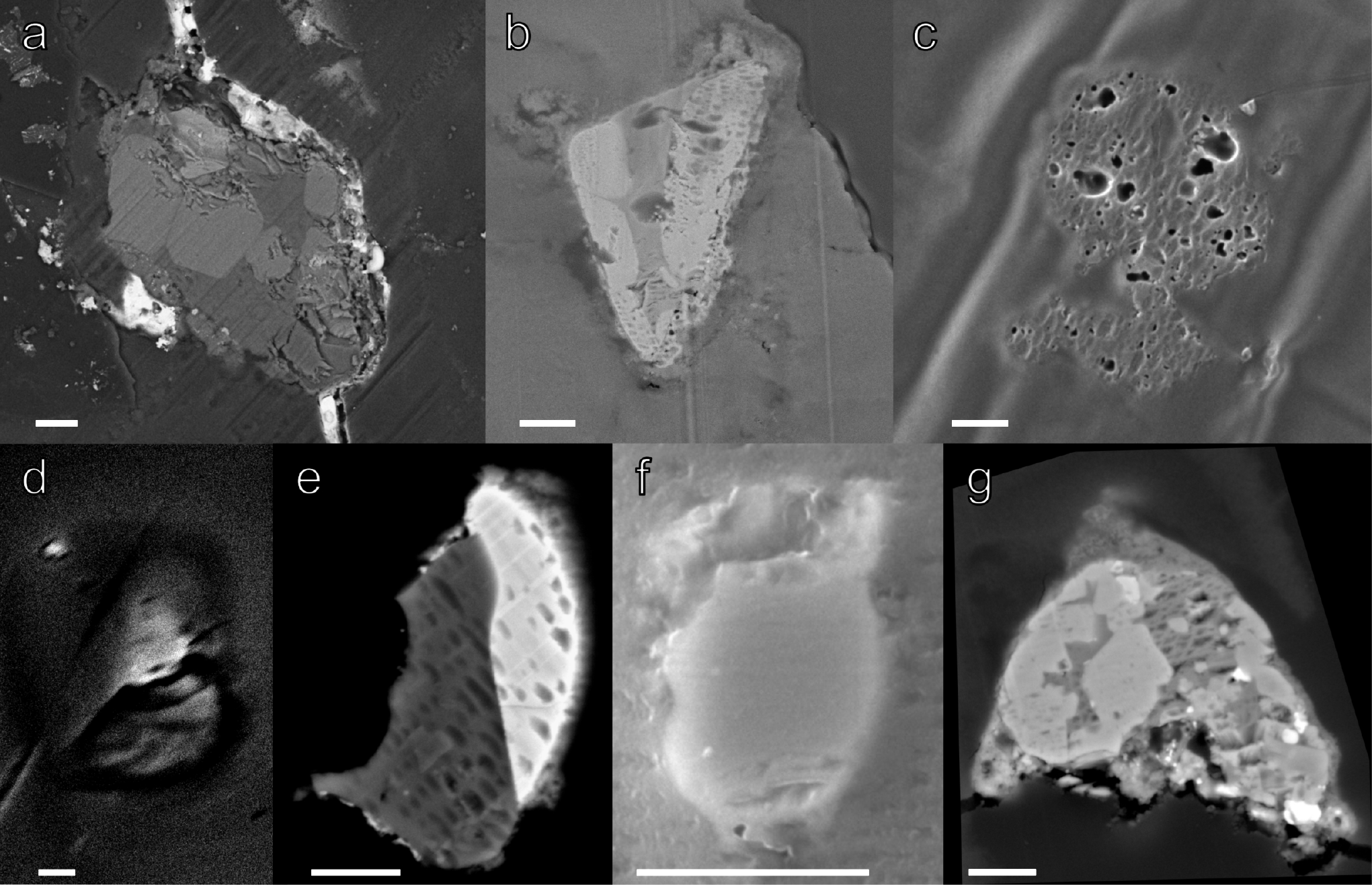}
\end{center}
\vspace*{-0.25in}
\caption{Electron microscope images (SE=secondary electrons, BSE=backscattered electrons) of seven Stardust fragments measured individually. White scale bar is 2 $\mu$m in all panels. a) Iris (BSE), b) Callie (BSE), c) Tintin (SE), d) Pooka (BSE), e) Caligula (BSE), f) The Butterfly (SE), g) Cecil (BSE).\label{largegrainfig}}
\end{figure}

\begin{figure}[!ht]
\begin{center}
\includegraphics[width=\textwidth]{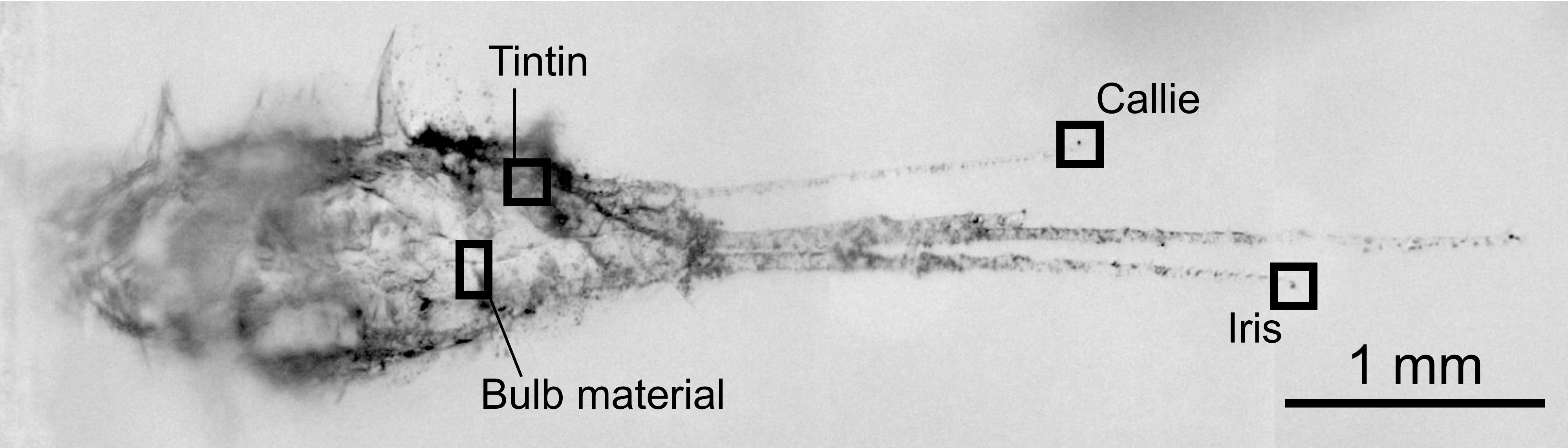}
\end{center}
\vspace*{-0.25in}
\caption{Photomicrograph of Stardust track C2052,74. Positions of Iris, Callie, Tintin (approximate), and the bulb material (approximate) are indicated.\label{track74fig}}
\end{figure}

%\subsubsection{Iris} 
\textbf{Iris: }Iris is a type-II chondrule fragment from Track C2052,74, 23$\times$10 $\mu$m in size, described in detail in \citet{Ogliore:2012p5583}. This particle contains Fe-rich olivines, chromite, and plagioclase in a SiO$_2$-rich, mostly glassy mesostasis.

%\subsubsection{Callie} 
\textbf{Callie: }Callie is a type-II chondrule fragment from Track C2052,74, 10$\times$6 $\mu$m in size, possibly related to Iris, described in detail in \citet{stodolna2013iron}. Like Iris, Callie contains subgrains of Fe-rich olivine, plagioclase, and chromite in a subcrystalline mesostasis.

%\subsubsection{Tintin} 
\textbf{Tintin: }Tintin is a 10$\times$8 $\mu$m  amorphous Fe-rich, Al-bearing silicate particle with a vesiculated texture from the bulb of track C2052,74.  

%\subsubsection{Pooka} 
\textbf{Pooka: }Pooka is a 3$\times$3 $\mu$m particle from track C2061,113 and consists mostly of pyroxene (En$_{80}$).

%\subsubsection{Caligula} 
\textbf{Caligula: }Caligula is the largest terminal particle in track C2035,105. It is a 12$\times$6 $\mu$m assemblage of pyrrhotite and amorphous silicate, with inclusions of pyroxene and sulfides \citep{gainsforth2013caligula}.

%\subsubsection{The Butterfly} 
\textbf{The Butterfly: }The Butterfly is a 3$\times$2 $\mu$m particle from the bulb region of track C2009,77. It consists mostly of pyroxene (En$_{88}$), with sub-$\mu$m Al- and Na-rich phases.

%\subsubsection{Cecil} 
\textbf{Cecil: }Cecil is a terminal particle from track C2062,162. It is  10$\times$8 $\mu$m in size and is composed of Fe-rich olivine (Fo$_{66}$, Fo$_{68}$), Ni-rich and Ni-poor sulfides, spinel, and pyroxene, with an amorphous silicate groundmass \citep{gainsforth2014ni}. 

\subsection{Bulb of Track 74}
\label{fgdescrip}
\begin{figure}[!ht]
       \centering
        \includegraphics[width=0.75\textwidth]{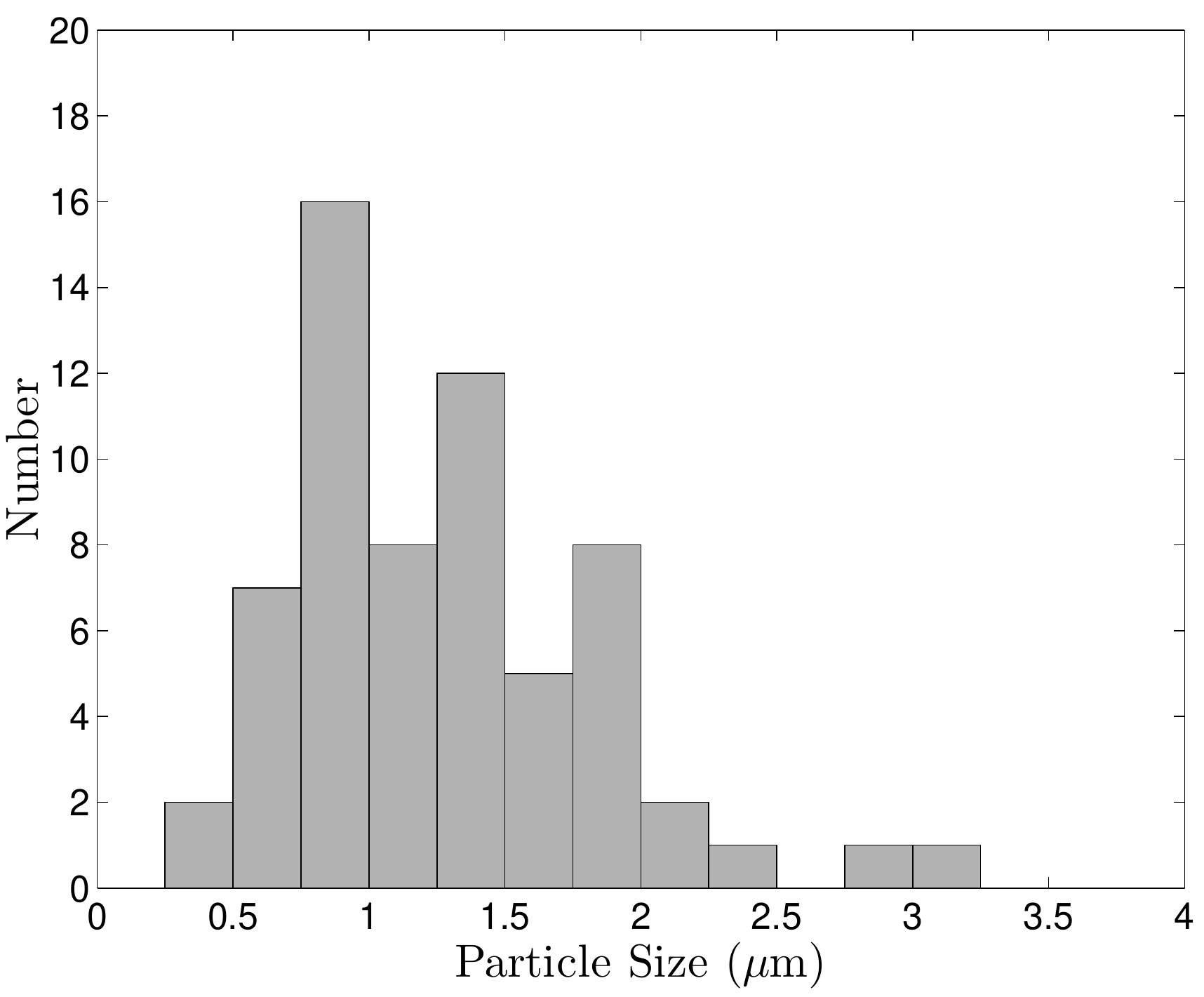}
        \caption{Size distribution (equivalent diameter) of the 63 measured Stardust fragments embedded in aerogel, extracted from the bulb of track C2052,74. Sizes were calculated from Mg-, Al-, or Fe-rich regions in the five scanning ion images shown in Figure \ref{siimaps}.\label{aerogel_sizes}}
 \end{figure}

We measured the O isotopic compositions of 63 particles from the bulb region of Stardust track C2052,74 (Figure \ref{track74fig}). This is a Type B track \citep[initial bulbous cavity tapering into one or more long, narrow tracks, see][]{burchell2008characteristics} with a total length of $\sim$6 mm. The widest part of the bulb is 2.5$\times$1 mm. The fine-grained material in the bulb was identified by its Mg-rich composition compared to the surrounding aerogel in an Energy Dispersive X-ray Spectroscopy (EDS) map. The size distribution of the 63 measured fragments is shown in Figure \ref{aerogel_sizes}. The Mg, Al, and Fe content of the fragments relative to O was measured by ion probe (Table \ref{finegrainedstats}). %, The Al, Mg, and Fe content of the fragments, as measured by ion probe, is given in Table \ref{finegrainedstats}.

\section{Sample Preparation}

%We prepared seven fragments from five different Stardust tracks for three-oxygen isotope analysis by SIMS, as well as a $\sim$1 mm$^2$ section of the bulb of Stardust track C2052,12,74 containing hundreds of $\mu$m-sized and smaller fragments. 

The individual fragments (Section \ref{coarsedescrip}) and Track 74 bulb particles (Section \ref{fgdescrip}) were prepared differently, as described below.

\subsection{Individual Fragments: Sample Preparation}
\label{grainmountsection}
Stardust tracks were extracted from their aerogel tiles as keystones \citep{Westphal:2010p5712} using glass needles robotically controlled by micromanipulators.  After synchrotron analyses  of fragments in the track \citep[X-ray fluorescence mapping and Fe K-edge X-ray absorption near-edge spectroscopy, see][]{Westphal:2009p1874}, the keystone is then dissected further to remove the fragment of interest in a thin ($\sim$100 $\mu$m) wafer of aerogel, which was embedded in epoxy and attached to the end of an epoxy bullet. Approximately 25 (volume) \% of the particle was ultramicrotomed into slices $\sim$100 nm thick and placed on transmission electron microscopy (TEM) grids with 10 nm amorphous carbon substrates, for analysis by TEM and synchrotron-based scanning transmission X-ray microscopy (STXM). The remaining $\sim$75\% of the particle remained in the epoxy bullet, its cross section exposed (a ``potted butt") for SIMS (secondary ion mass spectrometry) isotope measurements.

\begin{figure}[!ht]
       \centering
        \includegraphics[width=\textwidth]{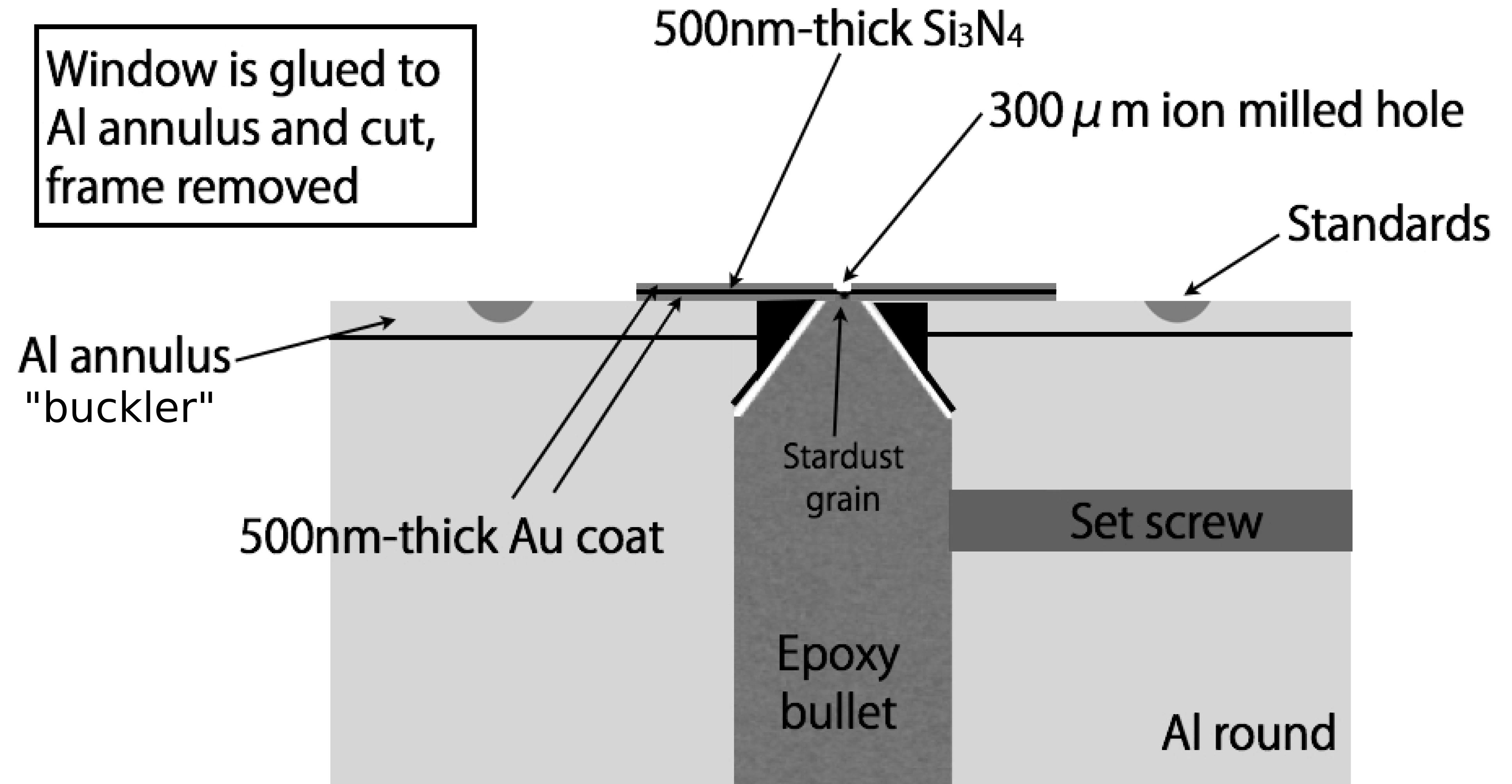}
        \caption{Reversible sample holder developed to make isotopic measurements of potted butts of Stardust particles by ion probe without remounting.\label{grainmount}}
        %  The potted butt of the Stardust grain in an epoxy bullet is first coated with Au and mounted in an Al holder.  Care is taken to make sure that the bullet is conductive and electrically connected to the aluminum holder.  The mount is covered by an aluminum plate called a ``buckler'', which has a gold-coated Si$_3$N$_4$ window with a 300$\mu$m hole in the center and a ring of standards mounted in epoxy and polished flat around the center.  The bullet is positioned vertically (Z) in the holder with a micromanipulator and the 300$\mu$m hole is positioned in X and Y so that the sample is exposed in the middle of the hole.  The entire mount is again C-coated to assure a continuous conductive coating over everything.\label{grainmount}}
 \end{figure}

The cylindrical potted butt was mounted in a holder specifically developed for SIMS analyses.  Our goal with this holder was to mount the potted butt without modification and make it possible to measure the isotopic compositions of the minerals exposed in the potted butt while retaining the ability to make additional microtome sections and to expose fresh surfaces for isotopic measurements \citep{Westphal:2011p5411}.  The mount consists of two components (Figure \ref{grainmount}): a Cameca ims 1280-compatible 1-inch aluminum round (2.54-cm diameter by 1-cm thick), which held the potted butt at a fixed altitude, and a ``buckler'', which simultaneously surrounds the particle with a flat conductive plane, matching the altitude of the particle, and integrates a ring of five unique mineral standards populating 12 positions around the sample. The multiple use of the five standards in the buckler allowed for the determination of any instrumental fractionation effects that depend on position in the buckler. No such instrumental fractionation effects were found.  The buckler consists of a thin Al disk with embedded standards, polished to a mirror finish, with a central 3~mm hole.  We mounted a Au-coated, 500~nm-thick Si$_3$N$_4$ membrane in which we had previously drilled a 300~$\mu$m hole using an ion mill, over the hole in the Al disk.  A motorized, encoded micromanipulator with 200nm resolution (Sutter MP-285) was used to integrate the complete assembly, since $\sim$1~$\mu$m precision is required in the three-dimensional alignment of the Si$_3$N$_4$ window with the particle. After assembly the sample was flat due to the ultramicrotomed face of the epoxy plateau except for a micron-sized step at the edge of the hole in the Si$_3$N$_4$ membrane more than 150~$\mu$m away from the Stardust particle. 

\subsection{Bulb of Track 74: Sample Preparation}
\label{bulbmaterialinaerogel}

\begin{figure}[!ht]
       \centering
        \includegraphics[width=\textwidth]{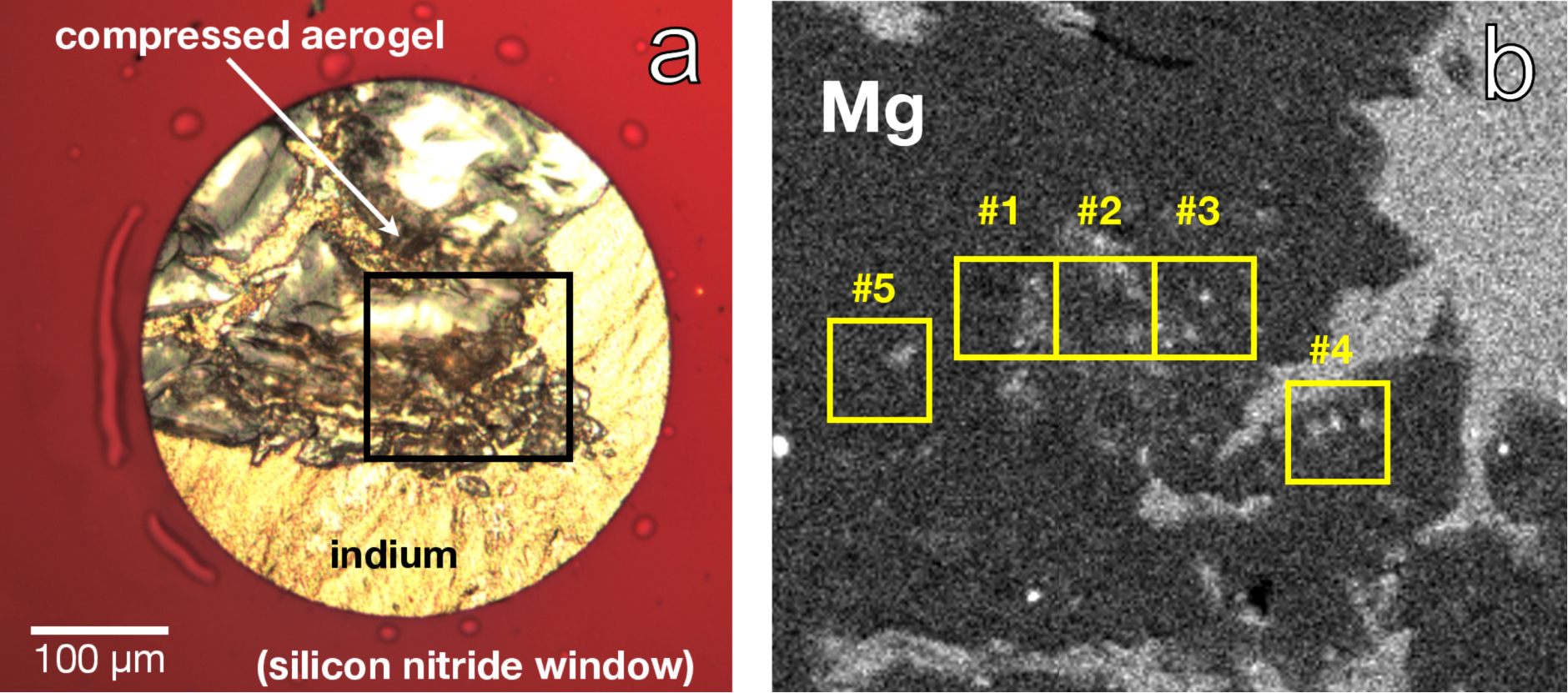}
        \caption{a) Compressed aerogel mounted for ion probe analysis. Area of Mg elemental map shown in b) is indicated by black rectangle. b) Mg elemental map of the region of compressed aerogel. The five 15$\times$15~$\mu$m analysis areas are outlined. False detection of Mg in indium surrounding the aerogel is due to the increased bremsstrahlung there.  \label{aerogelmount}}
 \end{figure}

We extracted a 1 mm$^2$ section from the wall of Stardust track C2052,74 using glass needles robotically controlled by micromanipulators. With a Teflon-coated anvil and an Instron press with a plunger calibrated for force, we compressed the aerogel sample into indium, which is pliable and electrically conductive. The aerogel porosity was sufficiently reduced to allow a conductive $\sim$50~nm Au coat to be applied. The samples were then mounted under an Au-coated Si$_3$N$_4$ window with a $\sim$400 $\mu$m ion-milled hole (Figure \ref{aerogelmount}a), similar to that described in Section \ref{grainmountsection}, to create a flat, conducting surface for SIMS analysis \citep{Westphal:2011p5411}.  We mapped the compressed aerogel using the University of Hawai`i's JEOL JXA-8500 field-emission gun electron probe microanalyzer to identify regions rich in Mg, which indicates the presence of cometary material (Figure \ref{aerogelmount}b). We identified five regions to analyze for O isotopes by SIMS. Additionally, we prepared two standards to assess the accuracy of our SIMS O isotope measurements: San Carlos olivine (shot into aerogel at $\sim$6 km/s, extracted as a keystone, pressed into indium) and spinel separates from the Allende meteorite (pressed into aerogel, then pressed into indium). The San Carlos olivine has O isotopic composition: \delox=5.25, \deloy=2.73; the Allende spinels have compositions that are well off the TFL: \deloxy $\approx -40\permil$ \citep{Makide:2009p6159}.

\section{Analytical Techniques}
The individual fragments (Section \ref{coarsedescrip}) and Track 74 bulb particles (Section \ref{fgdescrip}) were measured using different analytical conditions, as described below.
\subsection{Individual Fragments: Analytical Techniques}
\label{extractanalysiscond}
We measured three oxygen isotopes in the individually-mounted particles (Section \ref{grainmountsection}) using the University of Hawai`i Cameca ims 1280 ion microprobe in multicollection mode. A 25--30 pA Cs$^+$ primary ion beam was focused to $\sim$2 $\mu$m to allow for analysis of small phases in these particles. An electron flood gun was used for charge compensation. The detector configuration is given in Table \ref{ionprobedetectors}.

% (\ce{^{16}O}$^-$ on a Faraday cup (10$^{11}$ $\Omega$ resistor), \ce{^{17}O}$^-$ and \ce{^{18}O}$^-$ on electron multipliers)

\begin{table}[!ht]
\centering
\begin{tabular}{lccc}
\toprule 
Isotope&Detector&Mass-Resolving Power&Notes\\
\midrule
\ce{^{16}O} & L1 FC & $\sim$2000&10$^{11}$ $\Omega$ resistor\\
\ce{^{17}O} & Monocollection EM & $\sim$5500&\ce{^{16}OH^-} resolved from \ce{^{17}O^-}\\
\ce{^{18}O} & H2 EM & $\sim$2000&Exit Slit \#1 (500 $\mu$m)\\
\bottomrule
\end{tabular}
\caption{Detector configurations (UH Cameca ims 1280 ion probe) for measurements of individual comet Wild 2 fragments. EM = electron multiplier, FC = Faraday cup. \label{ionprobedetectors}}
\end{table}

Each measurement spot was presputtered for 3 minutes and then measured for 10 minutes (30 cycles, 20 seconds per cycle). The \ce{^{16}O} count rate for olivine was $\sim$3$\times$10$^7$ cps. The data was corrected for background, deadtime, and interference from \ce{^{16}OH-} to \ce{^{17}O-} (typically $<$0.2$\permil$). Variable yields between the Mono EM (\ce{^{17}O}) and H2 EM (\ce{^{18}O}) as well as mass-dependent isotope fractionation between the measured and true value of the terrestrial standards were simultaneously corrected by dividing the measured EM counts by a yield factor. The yield was simply the error-weighted mean of measured O isotope ratios (\ce{^{17}O}/\ce{^{16}O} for the Mono EM, \ce{^{18}O}/\ce{^{16}O} for the H2 EM) of the standard divided by the true ratio of the standard.

% \begin{eqnarray}
% Y_{Mono} = \frac{\textnormal{\deloy}_{\textnormal{measured}}+1000}{\textnormal{\deloy}_{\textnormal{true}}+1000}\\
% Y_{H2} = \frac{\textnormal{\delox}_{\textnormal{measured}}+1000}{\textnormal{\delox}_{\textnormal{true}}+1000}
% \end{eqnarray}

% where \greektext d\latintext \ce{^{17,18}O}$_{\textnormal{measured}}$ is the error-weighted mean of the standard measurements (calculated with Y=1) and \greektext d\latintext \ce{^{17,18}O}$_{\textnormal{true}}$ are the true \deloxy values of the standards.

The measured compositions of each mineral phase were corrected for instrumental mass fractionation by comparing with a set of measurements of appropriate mineral standards of known composition, made before and after the unknown (there was no significant drift in the standard measurements over time). If there were no appropriate standards available, we used San Carlos olivine to standardize the measurements: Caligula (amorphous silicate), Tintin (amorphous ferromagnesian silicate), and the mesostasis of Iris were standardized with San Carlos olivine. The magnitude of unaccounted instrumental mass fractionation for these measurements is likely no more than a few $\permil$, as their chemical compositions do not differ dramatically from olivine \citep[e.g., ][]{jogo2012heavily}, except for Tintin which is discussed in Section \ref{cgdiscussion}.

After SIMS measurements, we made sure we cleanly measured our intended targets by imaging the Stardust particles by secondary electron microscopy (SEM). We discarded any measurement spots that did not cleanly hit a single phase or that overlapped with the epoxy embedding medium, except for our measurement of Pooka. The measurement of this small particle overlapped marginally with the surrounding epoxy, and there was not enough area left to make another measurement. The epoxy has an isotopically light O composition on the TFL (\delox $\approx -85\permil$, \deloy $\approx -45\permil$, measured relative to a known mineral standard by ion probe) and \ce{^{16}O} count rate that is $\sim$1\% of enstatite or olivine. We corrected Pooka's measured O isotope composition using a simple mixing calculation: 
\begin{equation}
\left(\frac{^{j}\text{O}}{^{16}\text{O}}\right)_{\text{true}} = \left(\frac{^{j}\text{O}}{^{16}\text{O}}\right)_{\text{meas.}} +\left(X\right) \left(\frac{1-F}{F}\right) \left( \left(\frac{^{j}\text{O}}{^{16}\text{O}}\right)_{\text{meas.}} - \left(\frac{^{j}\text{O}}{^{16}\text{O}}\right)_{\text{epoxy}} \right)   
\label{mixingequation}
\end{equation}
where $X$ is the \ce{^{16}O} count rate in the epoxy divided by the count rate in the standard ($\text{C}_{\text{epoxy}}$/$\text{C}_{\text{stand.}}$), $j$ is the numerator isotope (i.e., 17 or 18), and $F$ is the volume fraction of the contaminated enstatite spot ($1-F$ was epoxy). It is possible to estimate F from the \ce{^{16}O} count rates: 

\begin{equation}
F = \frac{\text{C}_{\text{Pooka}} - \text{C}_{\text{stand.}}}{\text{C}_{\text{epoxy}} - \text{C}_{\text{stand.}}}
\label{mixingequationF}
\end{equation}

For Pooka, this correction resulted in a $+4.4\permil$ shift in \delox and $+2.3\permil$ shift in \deloy. The epoxy also affected the measurement by contributing a very large amount of \ce{^{16}OH-} to \ce{^{17}O-}. We measured the \ce{^{16}OH-} count rate after the measurement, and made a conservative assumption that the contribution to \ce{^{17}O-} had a 1$\sigma$ uncertainty of 5 ppm (mean value = 15 ppm). This correction resulted in a shift of $-2.4\permil$ in \deloy for Pooka. All other measurements had much smaller \ce{^{16}OH-} corrections ($<$0.2$\permil$), making the correction uncertainty negligible compared to the statistical and other systematic uncertainties.

% For grains large enough to make more than one SIMS measurements we report the weighted mean and its error. 

\subsection{Bulb of Track 74: Analytical Techniques}
Using a $<$3 pA primary Cs$^+$ beam on the University of Hawai`i Cameca ims 1280 ion probe focused to $\sim$500 nm, we acquired 128$\times$128 pixel, $15~\mu$m $\times$ $15~\mu$m scanning ion images for a total of $\sim$15 hours per map. In each $\sim$75 s measurement cycle, we collected \ce{^{16}O-}, \ce{^{17}O-}, \ce{^{18}O-} simultaneously on three electron multipliers (L1, Monocollection, H2) and then peak-jumped to \ce{^{28}Si-}, \ce{^{27}Al}, \ce{^{24}Mg}\ce{^{16}O-}, and \ce{^{56}Fe} (all on Monocollection EM). For two maps (\#4 and \#5) we collected \ce{^{27}Al}\ce{^{16}O-} and \ce{^{56}Fe}\ce{^{16}O-} instead of \ce{^{27}Al-} and \ce{^{56}Fe-} (because the oxides had a larger seconary ion yield) and also collected \ce{^{12}C-} and \ce{^{13}C-} (simultaneously) to search for presolar silicon carbide grains. The mass resolving power for \ce{^{17}O-} was $\sim$5500 to minimize the contribution of the \ce{^{16}OH-} interference on \ce{^{17}O-}. The electron flood gun was used for charge compensation. We registered the maps collected for each cycle using the \ce{^{16}O-} channel to account for position drift during the long measurement.

\begin{figure}[!ht]
\includegraphics[width=\columnwidth]{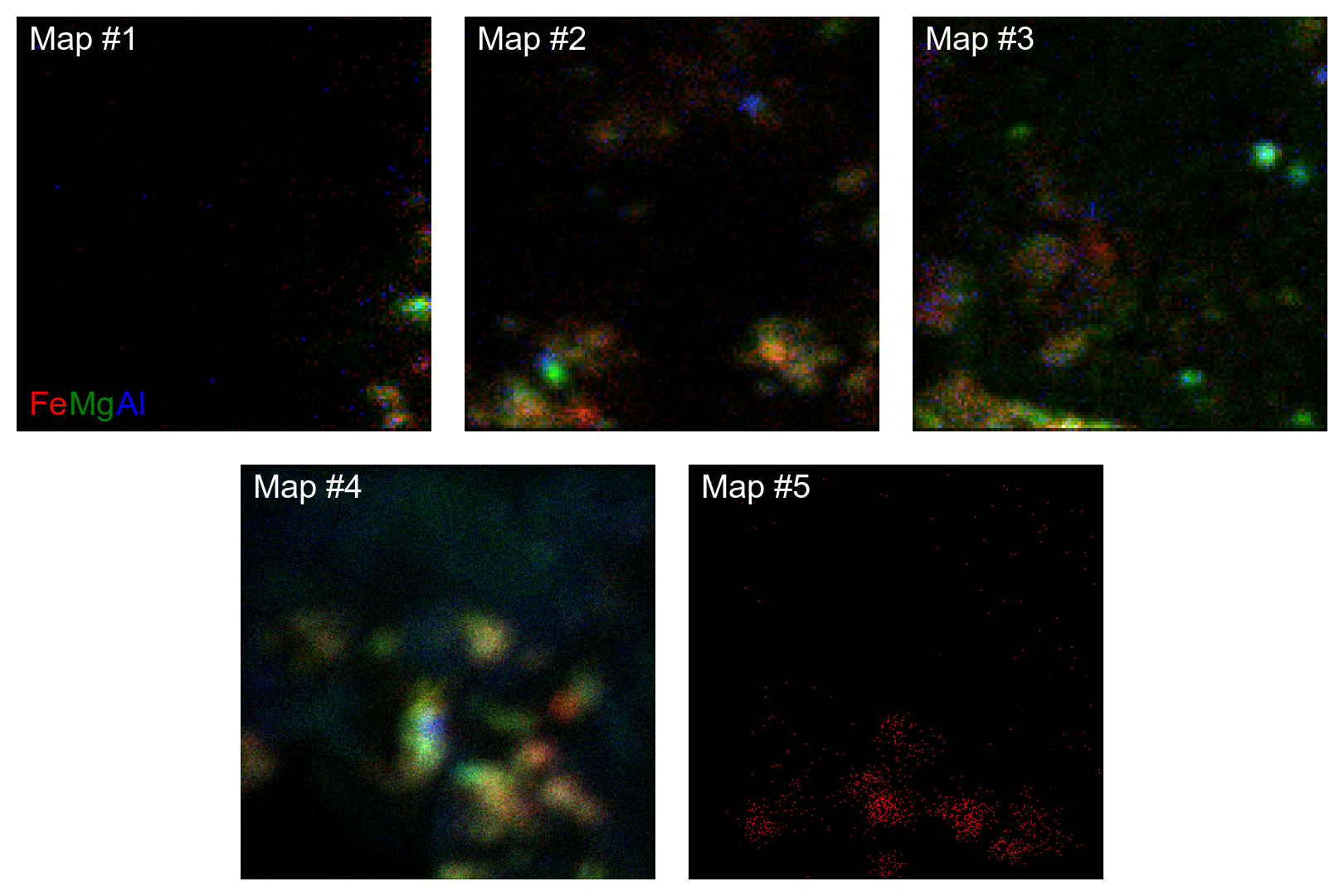}
\vspace{-2.2pc}
\caption{Scanning ion imaging maps of the five regions measured in the bulb of Stardust track C2052,74 measurements. Each map is 15$\times$15 $\mu$m. $^{56}$Fe$^{16}$O is red, $^{24}$Mg$^{16}$O is green, $^{27}$Al$^{16}$O is blue. The mass peaks for $^{24}$Mg$^{16}$O and $^{27}$Al$^{16}$O were lost during the measurement of Map \#5. \label{siimaps}}
\end{figure}

The goal of this measurement was to determine the O composition of the cometary (or standard) material relative to the surrounding aerogel, which has a known O isotopic composition, \delox$\approx$$-1.1\permil$ and \deloy$\approx$$-0.5\permil$ \citep{McKeegan:2006p936}. The cometary material (or standard) contains Mg, Al, or Fe at a much higher concentration than the aerogel. From our collected MgO, AlO, and FeO maps (Figure \ref{siimaps}) we were easily able to distinguish (using a simple thresholding algorithm or drawing by eye) spatially and compositionally distinct cometary particles from the surrounding aerogel.

We calculated \ce{^{18}O}/\ce{^{16}O} and \ce{^{17}O}/\ce{^{16}O} in the identified individual cometary particles. We normalized this composition to the surrounding aerogel O composition and calculated uncertainties by the following Monte Carlo method. For each region of interest consisting of $N$ cometary pixels, we randomly selected $N$ background aerogel pixels and calculated the \ce{^{18}O}/\ce{^{16}O} and \ce{^{17}O}/\ce{^{16}O} ratios in this set of aerogel pixels. We calculated \delox and \deloy for the region of interest relative to these aerogel ratios (while accounting for the known O composition of the aerogel), and repeated the process $10^5$ times. From this distribution, we computed a mean and standard deviation which correspond to the \delox and \deloy values and uncertainties for the cometary region. We also computed these values and uncertainties by selecting connected groups of $N$ aerogel pixels, instead of random unconnected pixels. The \delox and \deloy values were very similar in both analyses, but the uncertainties in the unconnected pixel analysis were larger. We chose to use the unconnected pixel analysis because it yielded more conservative errors. Analyzing a mixture of cometary material and aerogel would move the measured value closer to \deloxy~=~(0, 0). Each region of interest was chosen to minimize the contribution of surrounding aerogel.

\subsection{SEM, EPMA, FIB, TEM}
We used two SEMs with EDS for sample preparation and characterization: a JEOL 5900LV SEM (University of Hawai`i at M\={a}noa) with a Noran EDS system, and a Tescan Vega 3 with an Oxford EDS system (University of California, Berkeley). We used a JEOL JXA-8500F Electron Microprobe at the University of Hawai`i at M\={a}noa to acquire elemental maps in compressed aerogel from the bulb of track 74. The operating voltage was 15 keV, and we acquired maps of 15 elements (including Mg, Fe, O, Si, In) alternating on five wavelength-dispersive spectrometers.

We used a focused ion beam (FIB) instrument and a transmission electron microscope (TEM) to extract and analyze particles in compressed aerogel from the bulb of track 74. Both FIB and TEM were housed at the National Center for Electron Microscopy at Lawrence Berkeley National Laboratory. The FIB was an FEI Strata 235 Dual Beam and the TEM was a Zeiss Libra 200 MC operating at 200 keV. We used the TEM for energy-filtered imaging and electron diffraction.

\section{Results of Standard Measurements}
To assess the accuracy and precision of our ion probe measurements, we repeatedly measured standards of known O isotopic composition. Standards for individual fragments (Section \ref{coarsedescrip}) and Track 74 bulb particles (Section \ref{fgdescrip}) were measured differently, as described below.

\subsection{Individual Fragments: Results of Standard Measurements}

We measured a San Carlos olivine standard analog before and after our measurement of each particle of unknown O isotopic composition. This analog mount was prepared similar to the Stardust sample: a piece of San Carlos olivine was embedded in the tip of an epoxy bullet, microtomed, and mounted as described in Section \ref{grainmountsection}. The specifics of our sample preparation procedure evolved as we improved methods to, e.g., control the height of the epoxy bullet in the mount. For this reason, we used a contemporaneously prepared olivine analog for each measurement of individual Stardust particles. By comparing the measured composition of the center-mounted San Carlos olivine with the measured value of San Carlos olivine in the standard buckler, we could measure any systematic instrumental mass fractionation between the center-mounted particle and the buckler standard. Such a shift can be caused by slightly different sample topography between the center and buckler. Sample topography can significantly affect instrumental mass fractionation \citep{Kita:2009p5652}. For most of the measurements, this center-buckler shift is consistent with 0$\permil$ per amu, and is at most 1.5$\permil$ per amu. We constrained the center-buckler shift by measuring 5--10 spots on both the buckler and center of the San Carlos olivine analog sample before and after each unknown measurement. The different values between sessions likely reflect changes in sample topography from an evolving mounting procedure, so we correct each unknown value with the center-buckler shift in the contemporaneously mounted olivine standard, measured before and after the Stardust particle. There was no significant instrumental mass fractionation as a function of position within the center-mounted San Carlos olivine.

\subsection{Bulb of Track 74: Results of Standard Measurements}
\begin{figure}[htpb]
\includegraphics[width=\columnwidth]{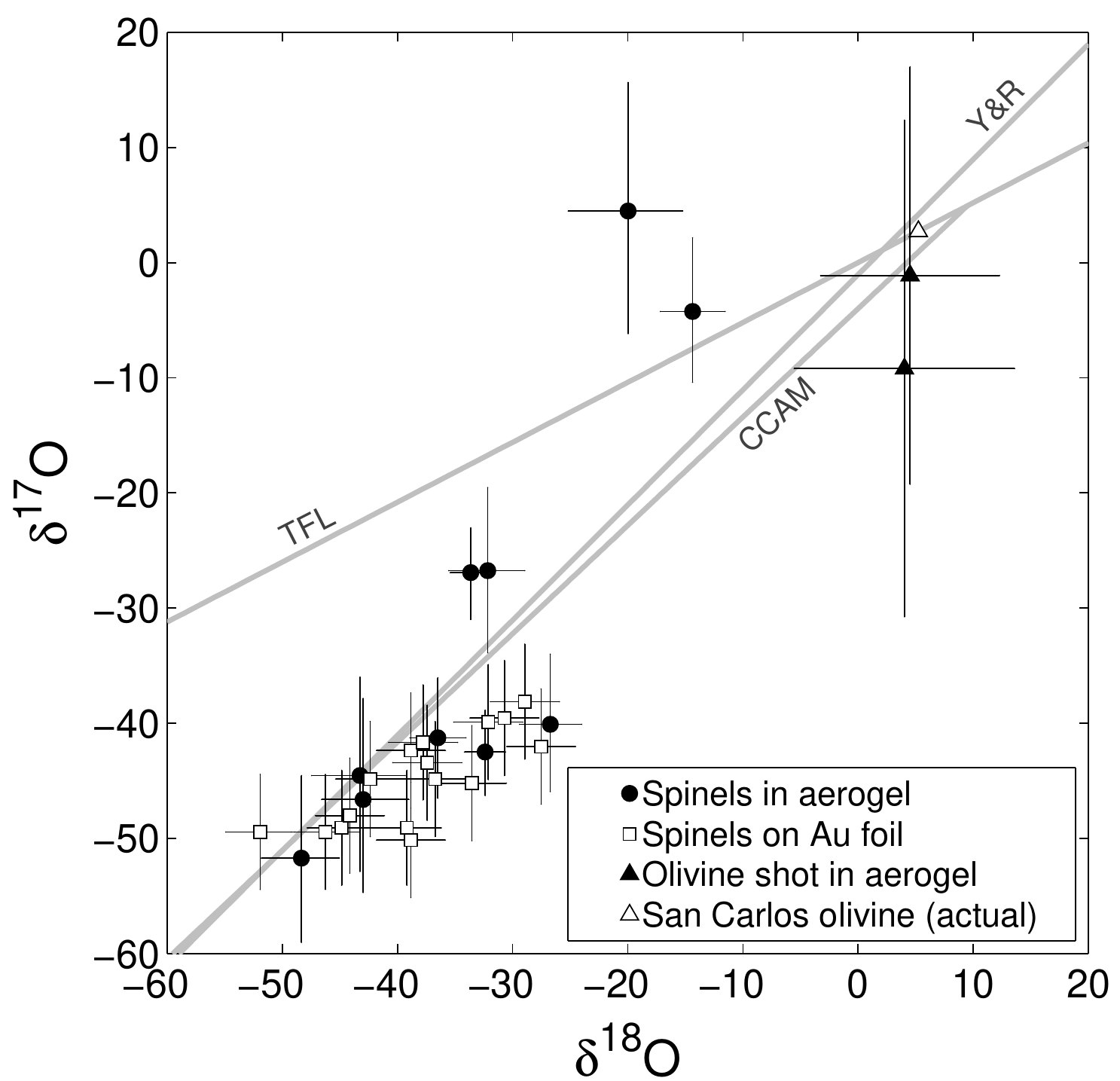}
%\vspace{-2.0pc}
\caption{Oxygen 3-isotope plot showing Allende spinel separates pressed into aerogel (black circles, this study), individual spinels measured on Au foil (open squares, \citet{Makide:2009p6159}), olivine shot into aerogel (black triangles, this study), and the true value of San Carlos olivine (open traingle). Uncertainties are 2$\sigma$. The terrestrial fraction line (TFL), carbonaceous chondrite anhydrous mineral line (CCAM, \citet{clayton1977distribution}) and Young \& Russell line (Y\&R, \citet{young1998oxygen}) are indicated. \label{fig:spinels}}
\end{figure}

We measured Allende spinels pressed into aerogel (Section \ref{bulbmaterialinaerogel}) to have O isotopic composition consistent with that measured for individual spinel grains distributed onto Au foil \citep{Makide:2009p6159}, except for two outliers (see Figure \ref{fig:spinels}). The two outliers measurements have cation/oxygen ratios (as measured by scanning ion imaging in the ion probe, Table \ref{finegrainedstats}) different from the other spots, indicating that they are non-spinel contaminants in the acid residue from Allende that were pressed alongside spinel grains into the aerogel. The precision of the pressed-aerogel spinel measurements was similar to that of spinels measured on Au foil. We measured the O isotopic composition of olivine shot into aerogel (Section \ref{bulbmaterialinaerogel}) to be consistent with its actual composition, though with large uncertainties due to the small size of fragments we measured. Matrix effects from analyzing minerals in aerogel are constrained to be small compared to the characteristic uncertainties of the measurements.

In order to estimate the Al/O, Mg/O, and Fe/O atom ratio of the Stardust fragments from ion probe measurements, we must constrain the relative sensitivity of these elements in the ion probe. It is well-known that different elements are detected with different efficiencies in the ion probe. The relative sensitivity factor (RSF) for two elements is defined as the ion probe elemental ratio divided by the ``true'' elemental ratio (the latter is typically measured by electron-probe microanalyzer (EPMA)). The RSF depends on a number of variables such as the primary ion species and the chemical composition of the substrate \citep[e.g.][]{bottazzi1992sims}. We measured an oxide (chromite) and silicate (pyroxene) of known chemical compositions using analytical conditions similar to our Stardust measurements. The mineralogy of the fragments we measure in compressed aerogel from the bulb of Track 74 is not known. However, the majority of the O-bearing fragments are likely silicates or oxides \citep{Stodolna:2012p5651}. The RSF can vary significantly between minerals (up to $\sim$2$\times$), so these estimates of Al/O, Mg/O, and Fe/O in the Stardust fragments have large uncertainties. The RSFs of Al/O, Mg/O, and Fe/O as calculated from ion probe and EPMA measurements of pyroxene and chromite standards, that are used in our estimates of these ratios in our Stardust measurements (Table \ref{finegrainedstats}), are given in Table \ref{RSFtable}.

\begin{table}[!ht]
\centering
\begin{tabular}{lcc}
\toprule 
Species Ratio&RSF$_{\textnormal{ Pyroxene}}$&RSF$_{\textnormal{ Chromite}}$\\
\midrule
\ce{^{27}Al}/\ce{^{16}O} & 0.015 & 6.1$\times$10$^{-3}$\\
\ce{^{27}Al^{16}O}/\ce{^{16}O} & 0.40 & 0.37\\
\ce{^{24}Mg^{16}O}/\ce{^{16}O} & 0.029 & 0.031\\
\ce{^{56}Fe}/\ce{^{16}O} & 6.0$\times$10$^{-3}$ & 4.0$\times$10$^{-3}$\\
\ce{^{56}Fe^{16}O}/\ce{^{16}O} & 0.024 & 0.045\\
\bottomrule
\end{tabular}
\caption{Relative sensitivity factors in pyroxene and chromite. \label{RSFtable}}
\end{table}

\section{O Isotopic Compositions of comet Wild 2 Particles}
\begin{figure}[!ht]
\begin{center}
\includegraphics[width=\textwidth]{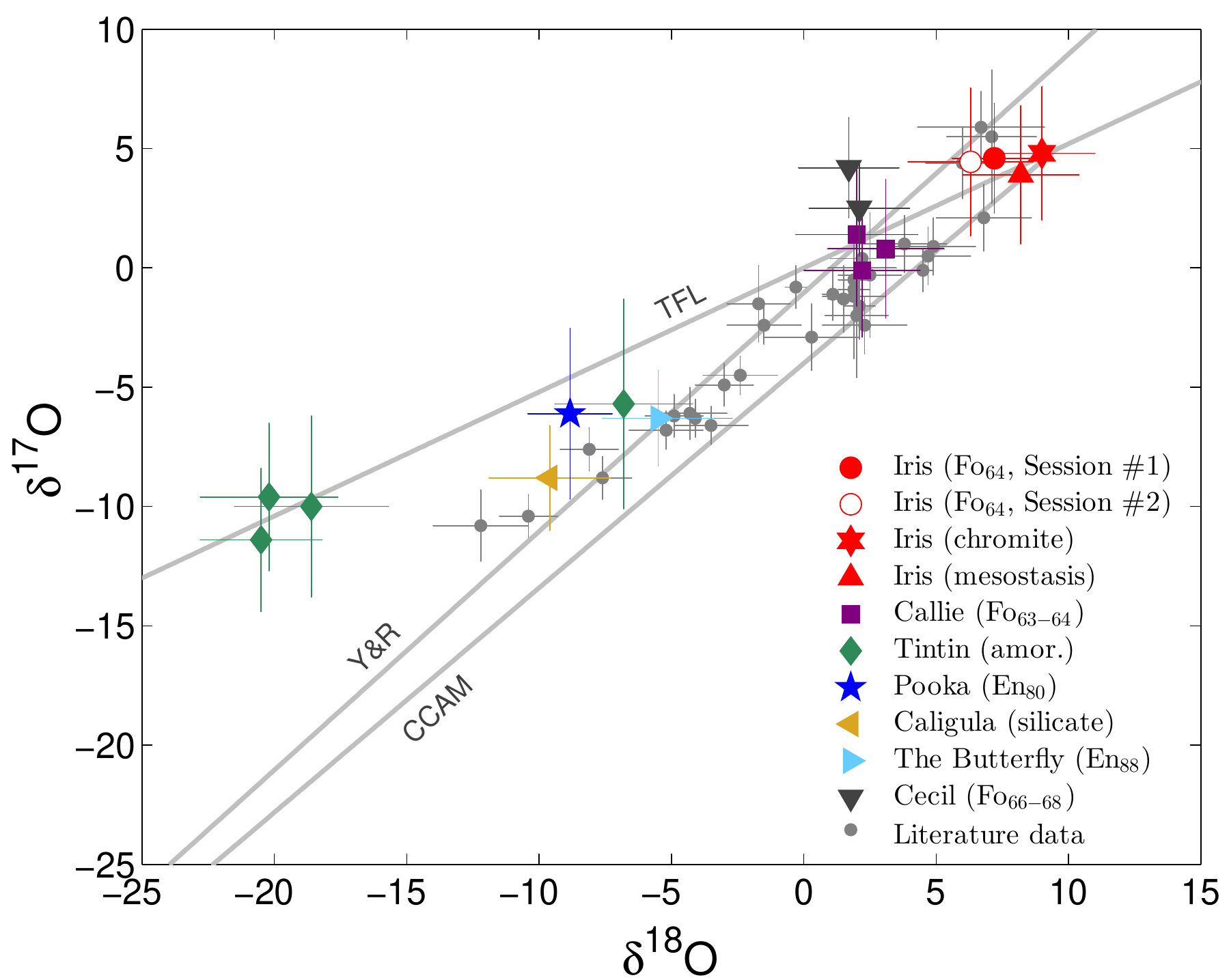}
\end{center}
\vspace*{-0.25in}
\caption{Oxygen 3-isotope plot showing the compositions of the seven extracted Wild 2 fragments. Iris olivine was measured in two separate sessions to verify our measurement reproducibility (Section \ref{reproducibilityextractedfragments}). Literature data are from \citet{McKeegan:2006p936,Nakamura:2008p2349,nakashima2012oxygen}; eight measurements of \ce{^{16}O}-rich particles from these are not shown. Uncertainties are 2$\sigma$. \label{StardustOfragments}}
%  \citet{young1998oxygen}
\end{figure}

Our measurements of the extracted Stardust fragments are shown in Figure \ref{StardustOfragments} and listed in Table \ref{StardustOfragmentstable}. Most of the measurements of the extracted $>$2~$\mu$m fragments are consistent with the terrestrial fractionation line.

\setlength{\tabcolsep}{6pt}
\begin{table}[!ht]
\small
\begin{tabular}{cccccLLL}
\hline  \TT \BB
Name&Size ($\mu$m)&Track&Location&Phase&\multicolumn{1}{c}{\delox\:($\permil$)}&\multicolumn{1}{c}{\deloy\:($\permil$)}&\multicolumn{1}{c}{\Deloy\:($\permil$)} \\
\hline\\[0.5ex]
%Iris& 23$\times$10& C2052,12,74 & Terminal & Fo$_{64}$ & 7.2\pm1.6 & 4.6\pm2.3 & 0.9\pm2.3 \\
Iris$^{\dagger}$& 23$\times$10& C2052,74 & Terminal & Fo$_{64}$ (1$^{\ddagger}$) & 7.2\pm1.6 & 4.6\pm2.3 & 0.9\pm2.4 \\
& &  &  & Fo$_{64}$ (2$^{\ddagger}$) & 6.3\pm2.4 & 4.4\pm3.1 & 1.2\pm3.3\\
& &  &  & chromite & 9.0\pm2.0 & 4.8\pm2.8 & 0.1\pm3.0\\
& &  &  & mesostasis & 8.2\pm2.2 & 3.9\pm2.9  & -0.4\pm3.1\\ [1ex]
%Callie& 10$\times$6 & C2052,12,74 & Terminal & Fo$_{64}$ (1) & 3.1\pm2.2 & 0.8\pm2.9&-0.8\pm3.1 \\
Callie& 10$\times$6 & C2052,74 & Terminal & Fo$_{64}$ (1) & 3.1\pm2.2 & 0.8\pm2.9&-0.8\pm3.1 \\
& & & & Fo$_{64}$ (2) & 2.2\pm2.2 & -0.1\pm2.8&-1.3\pm3.0 \\
& & & & Fo$_{63}$ & 2.0\pm2.3 & 1.4\pm3.0&0.4\pm3.2 \\
& & & & \emph{average} & 2.4\pm1.3 & 0.7\pm1.7 &  -0.6\pm1.8 \\ [1ex]
%Tintin & 10$\times$8 & C2052,12,74 & Bulb & amor. (1) & -18.6\pm2.9 & -10.0\pm3.8 & -0.3\pm4.1 \\
Tintin & 10$\times$8 & C2052,74 & Bulb & amor. (1) & -18.6\pm2.9 & -10.0\pm3.8 & -0.3\pm4.1 \\
& & & & amor. (2) &  -20.2\pm2.6 & -9.6\pm3.1 & 0.9\pm3.3 \\
& & & & amor. (3) & -6.8\pm2.6 & -5.7\pm4.4 & -2.1\pm4.6 \\
& & & & amor. (4) & -20.5\pm2.3 & -11.4\pm3.0 & -0.7\pm3.2 \\ [1ex]
%Pooka & 3$\times$3 & C2067,1,113 & Terminal & En$_{80}$ & -13.2\pm1.6 & -6.0\pm2.3 & 0.9\pm2.4 \\[1ex]
Pooka & 3$\times$3 & C2061,113 & Terminal & En$_{80}$ & -8.8\pm1.6 & -6.1\pm3.6 & -1.5\pm3.7 \\[1ex]
%Caligula & 12$\times$6 & C2035,5,105  & Terminal & silicate  &  -9.6\pm2.3 & -8.8\pm2.2 &-3.8\pm2.5 \\[1ex]
Caligula & 12$\times$6 & C2035,105  & Terminal & silicate  &  -9.6\pm2.3 & -8.8\pm2.2 &-3.8\pm2.5 \\[1ex]
%The Butterfly & 3$\times$2 & C2009,37,77 & Bulb & En$_{88}$ & -5.5\pm2.1  & -6.3\pm2.0 & -3.8\pm2.2 \\[1ex]
The Butterfly & 3$\times$2 & C2009,77 & Bulb & En$_{88}$ & -5.5\pm2.1  & -6.3\pm2.0 & -3.4\pm2.2 \\[1ex]
%Cecil & 10$\times$8 & C2062,2,162 & Terminal & Fo$_{66}$ &  1.7\pm1.9 & 4.2\pm2.1 & 3.4\pm2.3 \\
Cecil & 10$\times$8 & C2062,162 & Terminal & Fo$_{66}$ &  1.7\pm1.9 & 4.2\pm2.1 & 3.4\pm2.3 \\
& & & & Fo$_{68}$ & 2.1\pm1.9 & 2.5\pm2.0 & 1.4\pm2.2\\
& & & & \emph{average} & 1.9\pm1.3 & 3.4\pm1.4  & 2.4\pm1.6   \\ [1ex]
\hline
\end{tabular}
\caption{O isotope compositions of individual Stardust cometary particles. Parentheses indicate multiple measurement spots on the same grain. Errors are 2$\sigma$. ``amor.'' = amorphous. $^{\ddagger}$Iris olivine was measured in two separate sessions to verify our measurement reproducibility (Section \ref{reproducibilityextractedfragments}). $^{\dagger}$These measurements were reported previously in \citet{Ogliore:2012p5583}. \label{StardustOfragmentstable}}
\end{table}

\begin{figure}[!ht]
\begin{center}
\includegraphics[width=\textwidth]{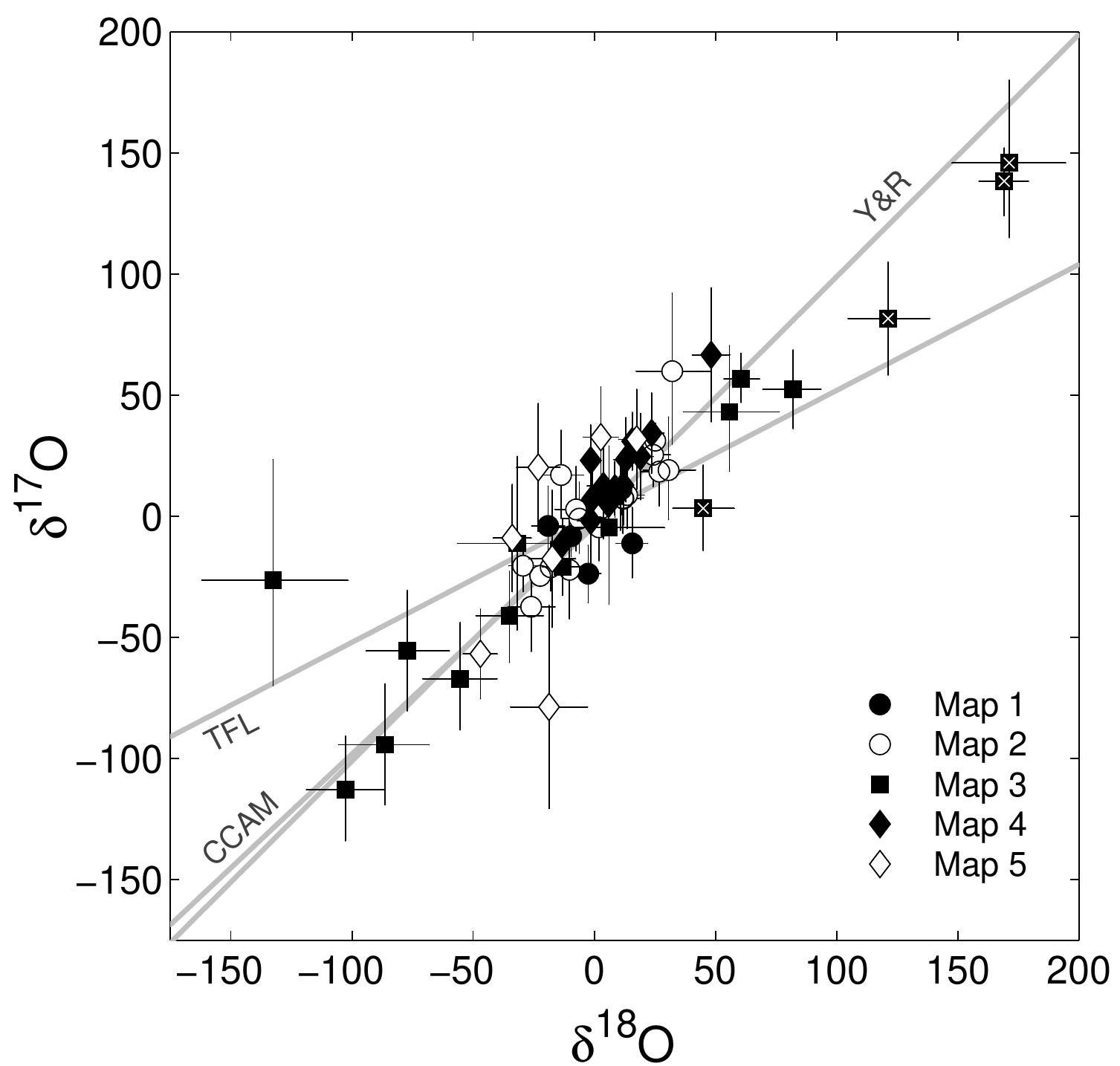}
\end{center}
\vspace*{-0.25in}
\caption{Oxygen 3-isotope plot showing the compositions of 63 particles in the bulb of Stardust track C2052,74. The terrestrial fraction line (TFL), carbonaceous chondrite anhydrous mineral line (CCAM) and Young \& Russell line (Y\&R) are indicated. Remeasured particles (Section \ref{remeasureaerogel}) are marked with a white X inside a black square. Uncertainties are 2$\sigma$. \label{StardustObulb}}
\end{figure}

Our O isotope measurements of the fine-grained material in the bulb of Stardust track C2052,74 are listed in Table \ref{finegrainedstats} and plotted in Figure \ref{StardustObulb}. The chemical composition (Mg, Al, and Fe relative to O) of these particles is also given in Table  \ref{finegrainedstats}. The Wild 2 fine-grained material shows a very broad range of O isotope compositions.

\clearpage
\newpage

\small{
\begin{center}
\begin{longtable}{>{$}l<{$} >{$}c<{$} >{$}c<{$} >{$}c<{$} >{$}c<{$} >{$}c<{$} >{$}c<{$} >{$}c<{$} >{$}c<{$}}
\hline
\text{Particle}&\text{Map}&\text{Size ($\mu$m)}&\text{Mg/O}&\text{Al/O}&\text{Fe/O}&\multicolumn{1}{c}{\text{\delox\:($\permil$)}}&\multicolumn{1}{c}{\text{\deloy\:($\permil$)}}&\multicolumn{1}{c}{\text{\Deloy\:($\permil$)}}\\
\hline
1 & 1 & 0.7 & 0.02 & <0.01 & 0.01 & 15.7\pm6.5 & -11.1\pm14.6 & -19.3\pm15.0\\
2 & 1 & 0.9 & 0.02 & <0.01 & 0.02 & -2.6\pm5.3 & -23.7\pm11.9 & -22.3\pm12.2\\
3 & 1 & 1.3 & 0.09 & <0.01 & 0.03 & -9.6\pm3.6 & -8.1\pm8.7 & -3.1\pm8.9\\
4 & 1 & 0.6 & 0.02 & <0.01 & 0.05 & -19.1\pm6.9 & -3.9\pm16.2 & 6.0\pm16.6\\
5 & 1 & 1.5 & 0.04 & <0.01 & 0.04 & -1.6\pm3.2 & 4.6\pm7.5 & 5.5\pm7.7\\
6 & 2 & 1.0 & 0.07 & <0.01 & 0.02 & 1.9\pm7.3 & -4.5\pm14.2 & -5.5\pm15.0\\
7 & 2 & 2.8 & 0.04 & <0.01 & 0.06 & 9.3\pm2.6 & 11.5\pm4.9 & 6.7\pm12.2\\
8 & 2 & 1.0 & 0.06 & <0.01 & 0.04 & 13.5\pm7.3 & 8.8\pm13.9 & 1.7\pm8.9\\
9 & 2 & 0.9 & 0.17 & 0.06 & 0.08 & 11.8\pm8.0 & 7.6\pm15.2 & 1.4\pm16.6\\
10 & 2 & 0.9 & 0.05 & <0.01 & 0.05 & 26.9\pm8.2 & 18.5\pm14.6 & 4.5\pm7.7\\
11 & 2 & 1.3 & 0.06 & <0.01 & 0.06 & 11.2\pm5.5 & 11.2\pm10.6 & 5.4\pm14.7\\
12 & 2 & 1.2 & 0.07 & <0.01 & 0.06 & -29.4\pm5.9 & -20.3\pm10.8 & -5.1\pm5.1\\
13 & 2 & 0.7 & 0.03 & <0.01 & 0.01 & 30.7\pm10.9 & 19.1\pm21.2 & 3.2\pm14.4\\
14 & 2 & 0.7 & 0.02 & <0.01 & 0.01 & -26.1\pm9.8 & -37.3\pm18.6 & -23.8\pm15.8\\
15 & 2 & 0.8 & 0.04 & <0.01 & 0.03 & -7.5\pm8.7 & 2.9\pm17.5 & 6.7\pm15.2\\
16 & 2 & 0.8 & 0.06 & <0.01 & 0.03 & -13.8\pm9.1 & 17.1\pm18.3 & 24.2\pm11.0\\
17 & 2 & 0.6 & 0.02 & <0.01 & 0.02 & -10.3\pm11.0 & -22.1\pm20.7 & -16.7\pm11.2\\
18 & 2 & 3.0 & 0.11 & <0.01 & 0.09 & -22.3\pm2.2 & -24.5\pm4.4 & -12.9\pm21.9\\
19 & 2 & 1.0 & 0.05 & <0.01 & 0.03 & 24.2\pm7.2 & 25.4\pm13.6 & 12.8\pm19.2\\
20 & 2 & 1.0 & 0.07 & <0.01 & 0.05 & -6.2\pm7.4 & -1.0\pm14.7 & 2.3\pm18.1\\
21 & 2 & 1.9 & 0.09 & <0.01 & 0.09 & 25.1\pm4.0 & 31.3\pm7.2 & 18.2\pm18.9\\
22 & 2 & 1.2 & 0.32 & 0.02 & 0.05 & 12.0\pm6.1 & 7.3\pm11.2 & 1.0\pm21.5\\
23 & 2 & 1.3 & 0.07 & <0.01 & 0.07 & -18.0\pm5.0 & -21.0\pm10.2 & -11.6\pm4.5\\
24 & 2 & 2.0 & 0.18 & 0.01 & 0.13 & 3.0\pm3.4 & 3.4\pm6.8 & 1.8\pm14.1\\
25 & 2 & 1.2 & 0.08 & <0.01 & 0.14 & 8.6\pm5.7 & 10.9\pm11.2 & 6.4\pm15.2\\
26 & 2 & 0.5 & 0.09 & <0.01 & 0.10 & 32.2\pm15.4 & 59.9\pm31.2 & 43.2\pm7.5\\
27 & 3 & 0.9 & 0.06 & <0.01 & 0.06 & -102.8\pm16.2 & -112.8\pm21.6 & -59.4\pm15.0\\
28 & 3 & 0.5 & 0.11 & 0.03 & 0.11 & -132.6\pm30.2 & -26.4\pm46.7 & 42.5\pm12.2\\
29 & 3 & 0.7 & 0.07 & <0.01 & 0.02 & 6.1\pm23.8 & -4.6\pm32.6 & -7.8\pm8.9\\
30 & 3 & 1.0 & 0.24 & 0.02 & 0.09 & -55.3\pm15.4 & -67.1\pm22.2 & -38.3\pm16.6\\
31 & 3 & 0.8 & 0.13 & 0.01 & 0.02 & -86.4\pm18.6 & -94.3\pm25.0 & -49.3\pm7.7\\
32 & 3 & 1.1 & 0.02 & <0.01 & 0.09 & -35.0\pm13.9 & -41.0\pm18.8 & -22.8\pm14.7\\
33 & 3 & 1.9 & 0.04 & <0.01 & 0.06 & -13.1\pm8.5 & -20.8\pm12.1 & -14.0\pm5.1\\
34 & 3 & 2.4 & 0.05 & 0.01 & 0.08 & 60.6\pm7.4 & 56.7\pm10.2 & 25.2\pm14.4\\
35 & 3 & 0.9 & 0.04 & <0.01 & 0.03 & -77.2\pm17.2 & -55.5\pm24.9 & -15.4\pm15.8\\
36 & 3 & 1.5 & 0.08 & <0.01 & 0.11 & 82.0\pm12.0 & 52.4\pm16.3 & 9.7\pm15.2\\
37 & 3 & 0.9 & 0.11 & 0.01 & 0.02 & 55.7\pm19.7 & 43.2\pm25.9 & 14.2\pm11.0\\
38^{\dagger} & 3 & 1.9 & 0.07 & <0.01 & 0.11 & 169.1\pm10.2 & 138.3\pm14.0 & 50.4\pm11.2\\
39^{\dagger} & 3 & 1.1 & 0.11 & <0.01 & 0.13 & 121.2\pm17.0 & 81.7\pm23.3 & 18.7\pm21.9\\
40^{\dagger} & 3 & 1.3 & 0.17 & <0.01 & 0.15 & 45.0\pm12.8 & 3.3\pm17.7 & -20.0\pm19.2\\
41^{\dagger} & 3 & 0.8 & 0.13 & <0.01 & 0.11 & 171.2\pm23.5 & 145.9\pm32.6 & 56.9\pm18.1\\
42 & 3 & 0.6 & 0.15 & <0.01 & 0.03 & -31.8\pm24.8 & -11.1\pm35.9 & 5.4\pm18.9\\
43 & 4 & 1.9 & 0.07 & <0.01 & 0.06 & 8.4\pm3.4 & 11.5\pm11.2 & 7.1\pm11.3\\
44 & 4 & 1.3 & 0.08 & <0.01 & 0.04 & 10.8\pm4.7 & 10.0\pm15.9 & 4.4\pm16.0\\
45 & 4 & 1.9 & 0.13 & 0.01 & 0.12 & 5.9\pm3.3 & 5.1\pm10.4 & 2.0\pm10.5\\
46 & 4 & 1.7 & 0.10 & <0.01 & 0.13 & 15.7\pm3.8 & 31.0\pm12.0 & 22.8\pm12.2\\
47 & 4 & 1.3 & 0.13 & <0.01 & 0.07 & 19.0\pm5.4 & 24.7\pm17.8 & 14.8\pm18.1\\
48 & 4 & 1.4 & 0.11 & <0.01 & 0.17 & -1.5\pm4.4 & 23.1\pm14.7 & 23.9\pm14.9\\
49 & 4 & 1.7 & 0.12 & 0.01 & 0.15 & -0.7\pm3.7 & 7.4\pm11.4 & 7.8\pm11.6\\
50 & 4 & 1.1 & 0.18 & 0.01 & 0.08 & -1.5\pm5.6 & -1.8\pm18.3 & -1.0\pm18.5\\
51 & 4 & 1.9 & 0.16 & 0.01 & 0.17 & -13.2\pm3.2 & -11.2\pm10.4 & -4.3\pm10.6\\
52 & 4 & 1.0 & 0.15 & <0.01 & 0.11 & 3.8\pm6.9 & 12.5\pm21.8 & 10.5\pm22.1\\
53 & 4 & 0.9 & 0.10 & <0.01 & 0.06 & 48.2\pm7.8 & 66.7\pm27.7 & 41.6\pm28.0\\
54 & 4 & 1.2 & 0.17 & <0.01 & 0.14 & 13.0\pm5.1 & 23.4\pm17.4 & 16.7\pm17.6\\
55 & 4 & 1.3 & 0.27 & 0.02 & 0.18 & 23.7\pm5.1 & 34.4\pm16.7 & 22.1\pm16.9\\
56 & 4 & 1.7 & 0.23 & 0.01 & 0.17 & 12.0\pm3.8 & 12.9\pm12.0 & 6.7\pm12.1\\
57 & 5 & 2.2 & \text{--} & \text{--} & 0.02 & -33.9\pm7.9 & -8.9\pm22.2 & 8.7\pm22.5\\
58 & 5 & 1.6 & \text{--} & \text{--} & 0.02 & 2.7\pm7.2 & 32.7\pm20.8 & 31.3\pm21.1\\
59 & 5 & 1.4 & \text{--} & \text{--} & 0.03 & -17.3\pm9.7 & -17.5\pm28.3 & -8.5\pm28.8\\
60 & 5 & 1.7 & \text{--} & \text{--} & 0.04 & 17.6\pm7.5 & 31.8\pm20.7 & 22.6\pm21.1\\
61 & 5 & 1.9 & \text{--} & \text{--} & 0.05 & -23.1\pm9.1 & 20.2\pm26.5 & 32.2\pm26.9\\
62 & 5 & 1.4 & \text{--} & \text{--} & 0.05 & -18.7\pm16.0 & -78.7\pm42.0 & -69.0\pm42.8\\
63 & 5 & 1.8 & \text{--} & \text{--} & 0.01 & -47.0\pm7.0 & -56.8\pm18.5 & -32.3\pm18.8\\
\hline
  \caption{The sizes, Mg/O, Al/O, Fe/O elemental (atomic) ratios, and oxygen isotopic compositions of 63 fragments from the bulb of Stardust track C2052,74.$^{\dagger}$These particles were removed by FIB and reanalyzed for O isotopes (see Section \ref{remeasureaerogel}). \label{finegrainedstats}}
\end{longtable}
\end{center}
  }

\clearpage

\subsection{Assessing the Reproducibility of O Isotope Measurements}

The accuracy of ion probe measurements can be reduced by various systematic errors caused by, for example, sample topography or conductivity, changes in the primary beam or secondary ion optics, or variable electron-multiplier efficiency. The magnitude of these systematic errors in ion probe measurements of O isotopes can be large compared to the $\permil$-level precision needed to address cosmochemical questions. It is therefore necessary to prove that systematic errors in ion probe measurements of small Stardust particles are small compared to the statistical error of the measurement and can safely be ignored.
Many types of systematic errors in ion probe analyses can be quantified by measurements of known and homogenous isotope standards which are prepared and mounted analogous to the unknown. Repeated measurements of standards before and after the unknown can be used to estimate systematic errors caused by, for example, drifts in electron-multiplier efficiencies. Sample topography and location as well as differences in charging between sample and standard, on the other hand, produce unpredictable electric-field variations, which can be significant and cannot be quantified with measurements of the standard during the measurement session. To quantify these systematic errors, it is necessary to measure the same sample in more than one measurement session. However, the small size of Stardust samples and difficult sample preparation makes this challenging.

\subsubsection{Individual Fragments: Reproducibility}
\label{reproducibilityextractedfragments}
\begin{figure}[!htpb]
\includegraphics[width=\columnwidth]{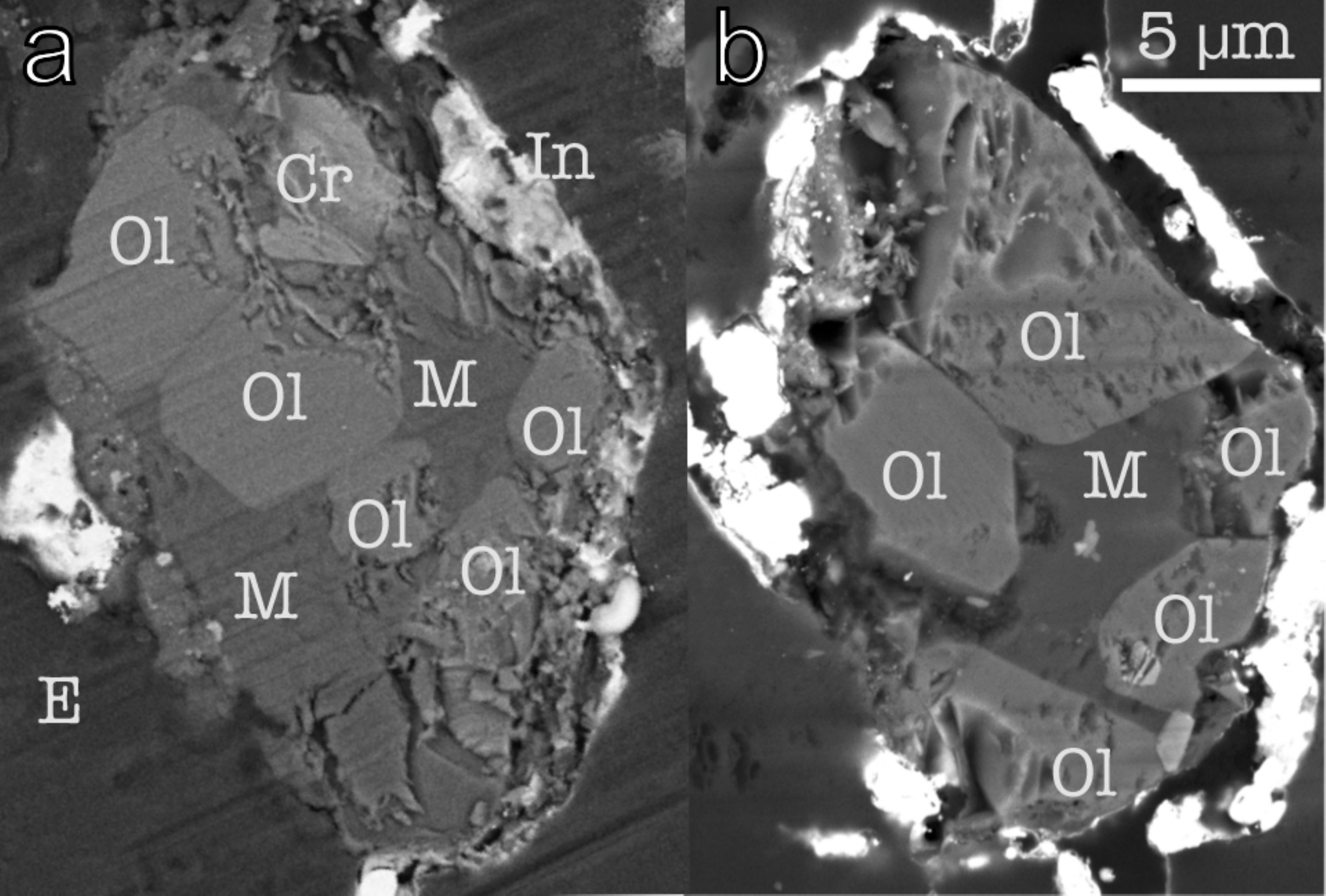}
\vspace{-1.9pc}
\caption{a) The exposed face of Iris in a potted butt before the first SIMS measurement of O isotopes. Ol=olivine, Cr=chromite, M=mesostasis, In=indium (embedding), E=epoxy (embedding). b) Iris after further microtoming, before the second SIMS measurement of O isotopes. \label{irisfig}}
\end{figure}

% \begin{figure}[!htpb]
% \includegraphics[width=\columnwidth]{plot_Iris_2013_comparison}
% \vspace{-2.2pc}
% \caption{O isotope measurements of olivine in Iris from Session \#1 (red squares, left image in Figure \ref{irisfig}) and Session \# 2 (blue squares, right image in Figure \ref{irisfig}). Error bars are 2$\sigma$. \label{irisO}}
% \end{figure}

% : 25--30 pA Cs$^{+}$ primary beam focused to a $\sim$2 $\mu$m spot, multicollection of O isotopes with $^{16}$O on a Faraday cup and $^{17}$O, $^{18}$O on electron multipliers, $\sim$5500 mass-resolving power for $^{17}$O$^-$ to minimize the contribution of the $^{16}$OH$^-$ interference on $^{17}$O$^-$. 

The type II chondrule fragment Iris (23$\times$10$\times$15~$\mu$m) from Stardust Track C2052,74 was microtomed ($\sim$100 nm thick slices), and the potted butt was remounted for SIMS analysis following the initial analyses \citep{Ogliore:2012p5583}. We measured the same olivine grains for O isotopes using analytical conditions similar to the first measurement (Section \ref{extractanalysiscond}). We corrected our measurement for any shift seen between the center and ring in an analogously prepared mount of San Carlos olivine. The microtomed faces of Iris before the first and second isotope measurement are shown in Figure \ref{irisfig}. The two faces have the same mineral assemblages with similar chemical compositions. Type II chondrules \citep[e.g.][]{jones1990petrology} can contain relict grains with chemical and isotopic compositions distinct from their host phenocrysts. No such grains appeared after microtoming further into Iris, so the O isotopic composition of the Iris olivine grains should be the same at these two different depths. 

The first O isotope measurement of an Iris olivine was \deloy=4.6$\pm$2.3$\permil$, \delox=7.2$\pm$1.6$\permil$ (2$\sigma$ uncertainties). Analyses of other olivines in Iris showed that they all have similar O composition. The second measurement of an olivine grain in the re-microtomed potted butt of Iris is \deloy=4.4$\pm$3.1$\permil$, \delox=6.3$\pm$2.4$\permil$, consistent with the first measurement made 2.5 years prior (Figure \ref{StardustOfragments}). Uncertainties were slightly larger in our second measurement due to slightly worse reproducibility of the standard measurements.

\subsubsection{Bulb of Track 74: Reproducibility}
\label{remeasureaerogel}
\begin{figure}[!ht]
\begin{center}
\includegraphics[width=\textwidth]{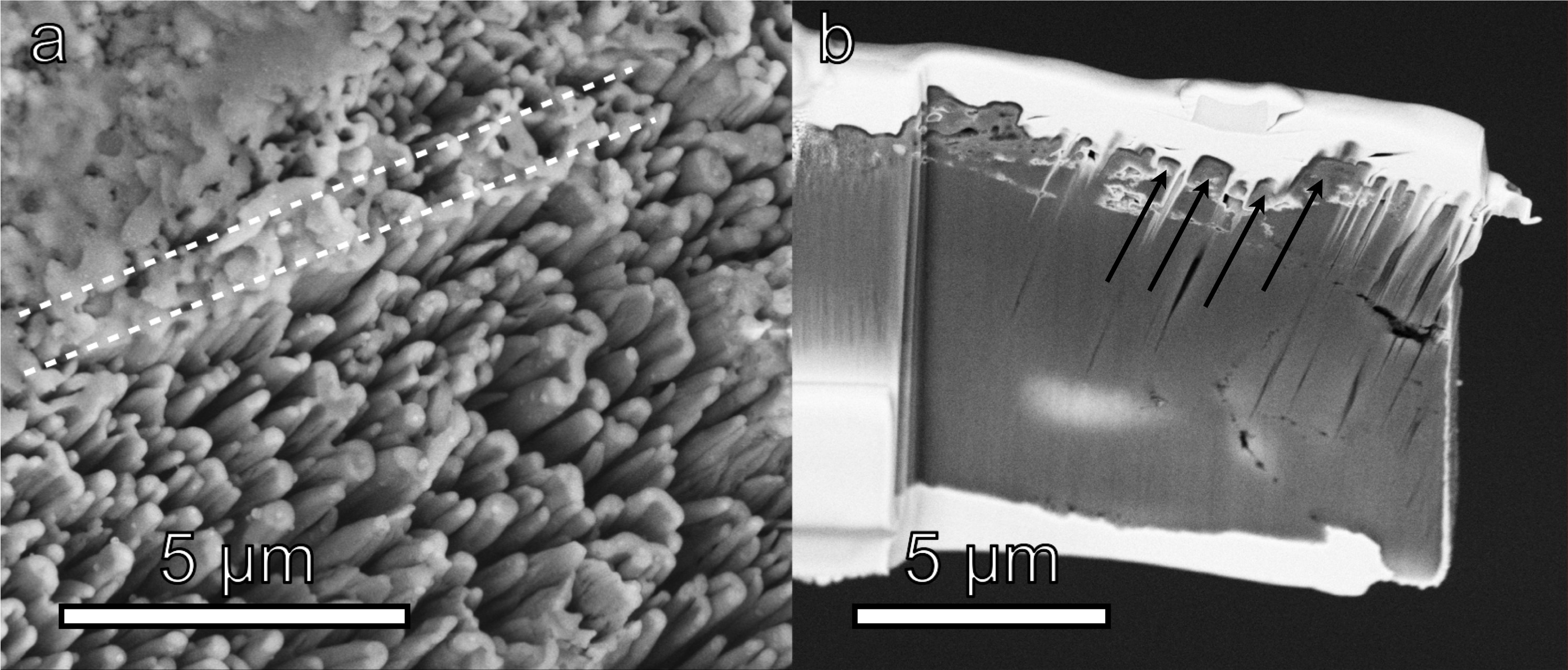}
\end{center}
\vspace*{-0.25in}
\caption{a) Secondary electron image of a region of compressed aerogel from the bulb of Stardust track C2052,74 after ion probe measurements. Dashed lines indicate the region of the FIB slice to remove the cometary material of interest. b) TEM bright-field image of the FIB slice after removal. White material at the top of the slice is deposited Pt, cometary particles are visible just below (four are indicated by black arrows). \label{rodanfib}}
\end{figure}

\vspace{-0.5pc}
\begin{figure}[!htpb]
\includegraphics[width=\columnwidth]{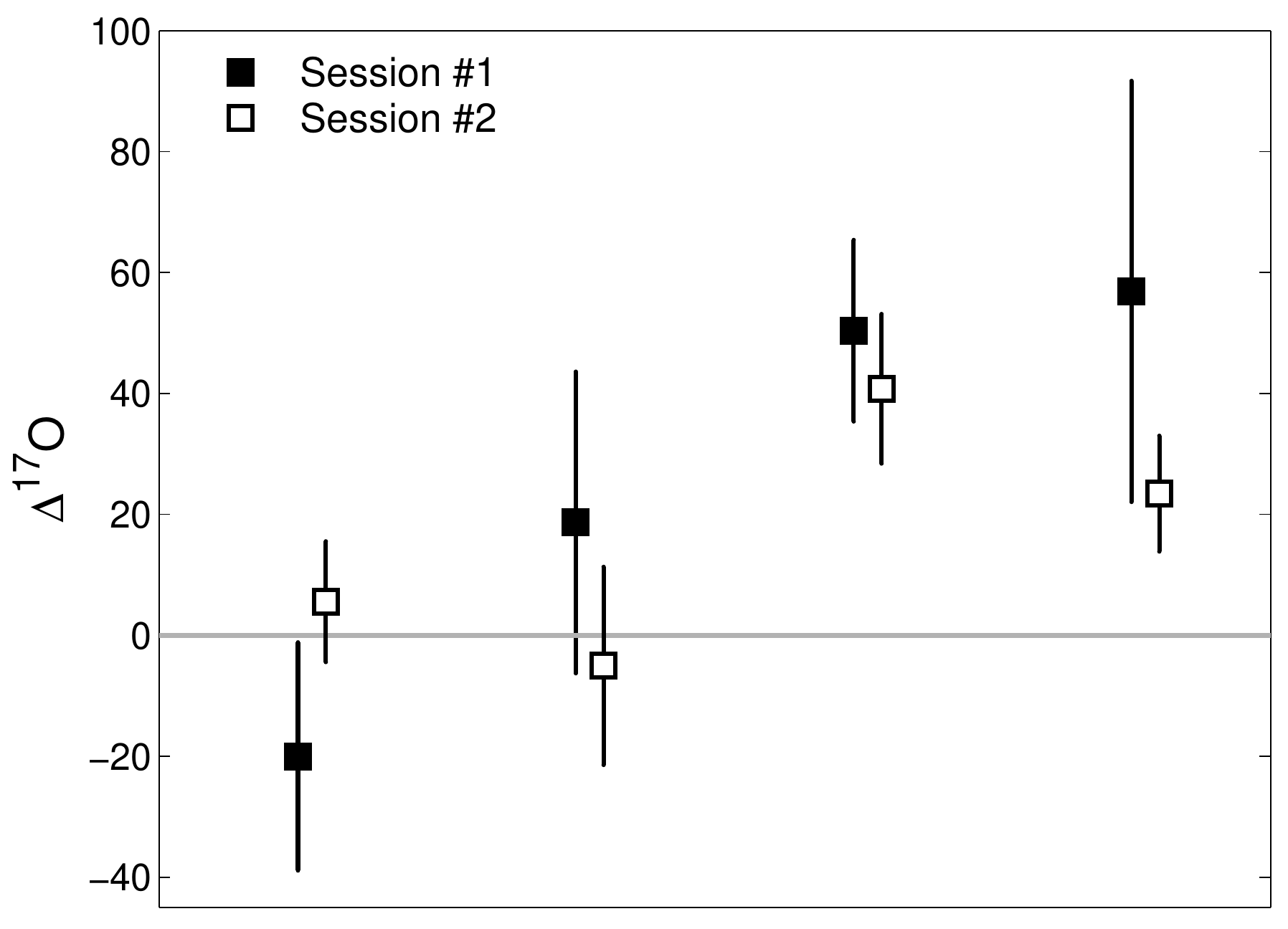}
%\vspace{-2.4pc}
\caption{$\Delta^{17}$O of four $\mu$m-sized particles from the wall of track C2052,74 made in compressed aerogel (filled squares: Session \#1) and in the same four particles in a removed FIB section (open squares: Session \#2). Uncertainties are 2$\sigma$. Session \#2 data are not included in Figure \ref{StardustObulb} and Table \ref{finegrainedstats}. \label{rodanO}}
\end{figure}

Two Fe-rich $\sim$1-$\mu$m particles in the wall of Stardust Track C2052,74 were measured to have high \deloxy (and \Deloy) compositions (Figure \ref{StardustObulb}). We removed a slice of the compressed aerogel containing these particles and others by FIB, and placed the lamella on Au foil, exposing the particles for additional ion probe analysis (Figure \ref{rodanfib}). We identified Fe-rich particles (1--2 $\mu$m in size) in the FIB lamella by SEM-EDS. The chemical compositions of four particles were qualitatively consistent with the previous ion probe measurements. The particles were identified by high-angle annular dark field imaging in the TEM and were found to be amorphous by electron diffraction. It is possible that the particles were originally crystalline but were amorphized by the SIMS analysis.

We obtained isotope maps of three oxygen isotopes, $^{16}$OH, and $^{56}$Fe$^{16}$O of cometary material in this FIB slice, using a $<$3 pA Cs$^{+}$ primary beam focused to $\sim$250 nm, slightly lower beam current and spot size than used in the previous measurement. We compared the \Deloy of four particles in the FIB slice with measurements of those four particles in the aerogel. Mass-dependent instrumental fractionation due to the different sample mounting and analytical conditions will affect the measured \deloxy of these four particles but should not affect \Deloy. The FIB slice measurements generally confirmed the results of our first measurement of the four particles in aerogel, as shown in Figure \ref{rodanO}. Discrepancies between these two measurements could be due to analyzing a different phase within a very small, multiphase particle, or effects from FIB sputtering or organometallic Pt deposition.

%The $^{16}$O-poor, Fe-rich region has O isotopic composition similar to cosmic symplectite \citep{seto2008mineralogical}. However, cosmic symplectite is made of magnetite and pentlandite and the $^{16}$O-poor Wild 2 material is amorphous. This Fe-rich material could have been an Fe-bearing crystalline object, such as magnetite, in the comet before being amorphized by high-speed capture. Our ion probe measurements of these grains could have also amorphized them prior to TEM analysis. It is thought that Acfer 094 cosmic symplectite formed by the oxidation and sulfurization of Fe metal grains through interaction with outer nebular water \citep{seto2008mineralogical}. The anomalously \ce{^{17,18}O}-rich Wild 2 material could have formed in a similar way before being incorporated into the comet.

%\clearpage
\section{Discussion}
\subsection{Coarse-grained material}
\label{cgdiscussion}
The spread of O isotopic compositions in the seven individual measured fragments is $\sim$19$\permil$ in \delox (excluding Tintin) with a standard deviation of $\sim$6$\permil$. These measurements, made on fragments from five different tracks, reflect the diversity generally seen in Stardust samples in other analyses of, e.g., the chemical composition of Wild 2 olivine \citep{frank2014olivine}. Our O isotope measurements (again, excluding Tintin) are similar to the range seen in previous \deloxy measurements of Wild 2 particles made by \citet{McKeegan:2006p936}, \citet{Nakamura:2008p2349}, and \citet{nakashima2012oxygen} (Figure \ref{StardustOfragments}). However, we did not measure any \ce{^{16}O}-rich grains \citep[e.g. \deloxy $\approx$ ($-40$); ][]{McKeegan:2006p936}. These grains, often low-iron and manganese-enriched (LIME) olivines, appear to be much less common than \ce{^{16}O}-poor Wild 2 grains, so this is not surprising.

Iris and Callie, two large terminal particles from Track C2052,74, are isotopically heavy, reflecting their likely origin as type-II chondrules \citep{Ogliore:2012p5583,gainsforth2015constraints}.

Cecil has similar O isotopic composition as Callie and is slightly lighter than Iris, but is from a different track. The igneous texture and mineralogy of portions of Cecil --- Fe-rich olivines in a glassy matrix, with Fe-rich pyroxene and sulfides --- shows some similarities to Iris and Callie. Further detailed study is required to know if some grains in Cecil are closely related to type II chondrule fragments, such as Iris and Callie, which appear to be common in comet Wild 2 (see also the Torajiro particle in \citet{Nakamura:2008p2349}). 

\begin{figure}[!ht]
\begin{center}
\includegraphics[width=\textwidth]{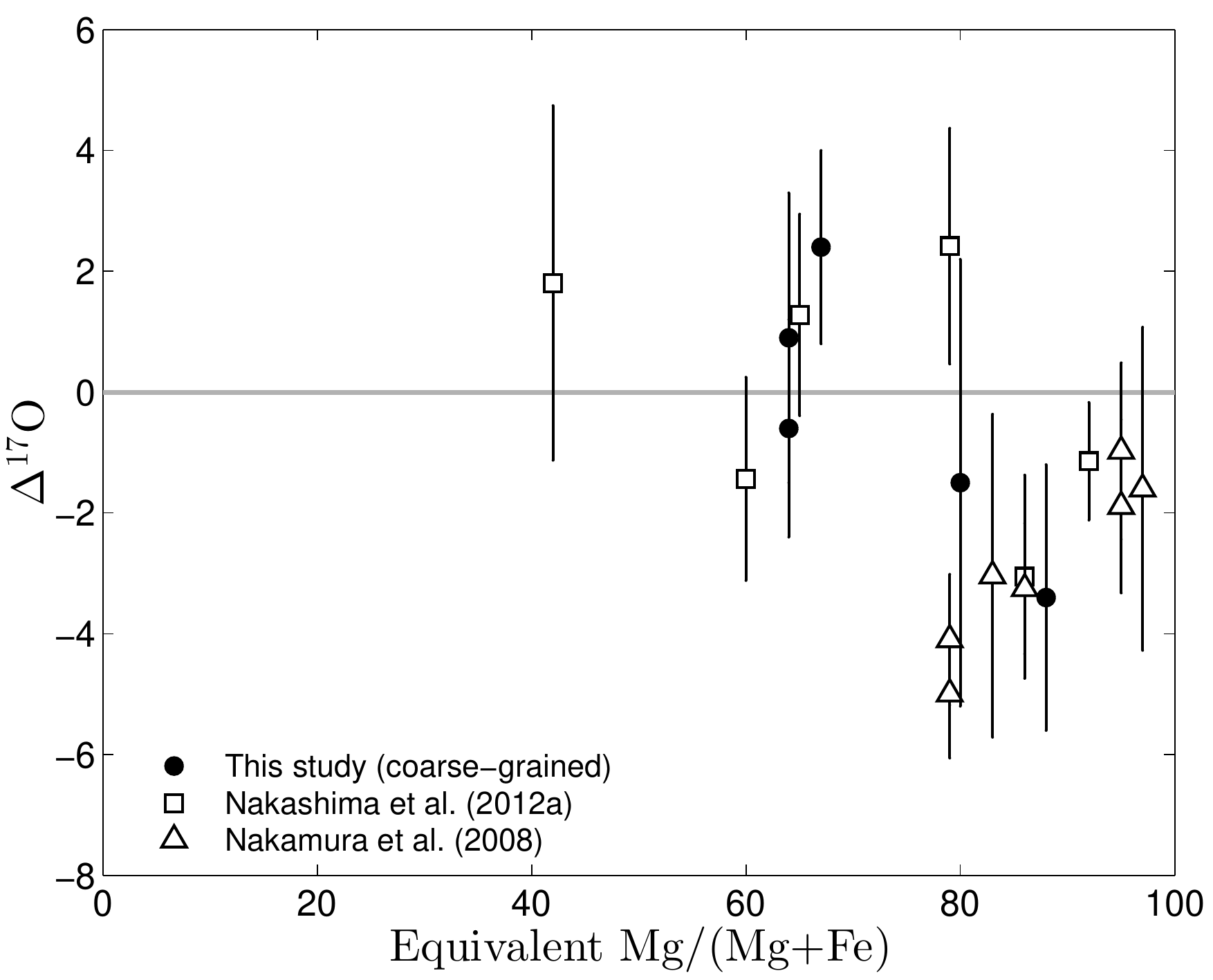}
\end{center}
\vspace*{-0.25in}
\caption{\Deloy vs. equivalent Mg/(Mg+Fe) (or, Mg\#) of Iris, Callie, Pooka, The Butterfly, and Cecil (filled circles) compared with literature data of other Wild 2 measurements, excluding $^{16}$O-rich particles (open squares: \citet{nakashima2012oxygen}, open triangles: \citet{Nakamura:2008p2349}). Error bars are 2$\sigma$. \label{fonumber_large_grains}}
\end{figure}

Pooka and The Butterfly are relatively simple pyroxene grains. Without the mineralogical context often seen in larger, multiphase Stardust terminal particles, it is difficult to know much about their origin. Both particles plot near the Young \& Russell line (Figure \ref{StardustOfragments}) with O isotopic compositions similar to Mg-rich chondrules \citep[e.g.][]{tenner2015oxygen}. Pooka and The Butterfly formed from a more $^{16}$O-rich reservoir than Cecil, Iris, and Callie. Ferromagnesian crystalline silicates from Wild 2 tend to show \Deloy values around $-2\permil$ for Fe-poor (Mg\# = 94--97) fragments and slightly higher values ($\sim$0$\permil$) for  more Fe-rich (Mg\#$<$85) fragments \citep{nakashima2012oxygen}. Only one of our particles had a Mg\#$>$85: The Butterfly, En$_{88}$. This particle had the lowest \Deloy measured in the crystalline phases of the extracted fragments: $-3.4\pm2.2\permil$. Grains in Iris, Callie, and Pooka (Mg\#=64, 64, and 80, respectively) had higher \Deloy values consistent with 0$\permil$. One Fo$_{66}$ grain in Cecil plots just above the TFL (\Deloy = $3.4\pm2.3\permil$). The \Deloy of our five particles with crystalline ferromagnesian phases as a function of Mg/(Mg+Fe) is plotted in Figure \ref{fonumber_large_grains}. Our measurements are consistent with the trend of \Deloy vs. Mg\# proposed by \citet{nakashima2012oxygen}.

% Our other crystalline ferromagnesian silicates, with more Fe-rich compositions, had higher \Deloy values consistent with 0$\permil$ (except for the Fo$_{66}$ grain in Cecil, which is close to being consistent with the TFL: \Deloy = $3.4\pm2.3\permil$). 

The amorphous silicate we measured in Caligula had O composition similar to Pooka and The Butterfly, indicating it formed from a similar O isotopic reservoir. 

Tintin is amorphous and shows a curious vesiculated texture, which could be the result of flash heating during aerogel capture. It has a heterogenous O isotopic composition --- one of the four measured spots was 13$\permil$ greater in \delox than the other three. We did not have an appropriate O isotope standard for the Fe-rich, Al-bearing silicate composition of Tintin. We used San Carlos olivine for the standard, which will result in a systematic uncertainty (due to matrix effects) of possibly up to a few $\permil$ parallel to the TFL. Three of the four spots we measured were more than 10$\permil$ isotopically light on the terrestrial fractionation line compared to the fourth spot. 

The epoxy embedding medium has an isotopically light composition so contamination with the epoxy can potentially explain our Tintin measurements. Secondary electron images of Tintin after ion probe measurements showed that the measurement spots did not overlap with the surrounding epoxy, but in principle we could have measured epoxy that penetrated the interior of this vesiculated particle. However, we can rule this out using the mixing calculation given in Equation \ref{mixingequation}. Because Tintin's count rate ($\sim$2.5$\times$10$^6$) was much closer to a mineral standard ($\sim$3.5$\times$10$^6$ for San Carlos olivine) than the epoxy ($\sim$2.5$\times$10$^4$), the contamination of the measured O isotopic composition from epoxy has to be minimal: at most $\sim$1$\permil$ in \delox and $\sim$0.5$\permil$ in \deloy.

%for the following reason. Epoxy has a much lower \ce{^{16}O} count rate than San Carlos olivine ($\sim$2.5$\times$10$^4$ for epoxy, $\sim$3.5$\times$10$^6$ for olivine using Tintin's measurement conditions). The \ce{^{16}O} count rate for Tintin was $\sim$2.5$\times$10$^6$. Based on a simple mixing calculation, this would only change the composition of Tintin by $\sim$1$\permil$ in \delox and $\sim$0.5$\permil$ in \deloy.

% It is possible that the vesiculated morphology of Tintin creates charging or electric-field distortions at the sample surface that result in a mass-dependent fractionation during the measurement. 

% Tintin's variable and light measured O composition could be the real composition of this particle (possibly altered from its pre-capture composition by heating/evaporative processes during aerogel capture), but we cannot rule out significant measurement artifacts.

Tintin's variable and light measured O composition is likely the real composition of this particle, possibly altered from its pre-capture composition by heating/evaporative processes during aerogel capture.

% Tintin was extracted from the bulb of a relatively large Stardust track, C2052,74.

% We conclude that a particle in the Track C2052,74 impactor was heated severely by high-speed capture, partially vaporized, then the vapor quickly recondensed to form Tintin. Tintin's light, mass-dependent O isotope fractionation is possibly the result of Rayleigh distillation which produces an isotopically light vapor and heavy residue. The very fast recondensation created the vesiculated texture and prevented the O isotopes from equilibrating in Tintin.

\subsection{Presolar grain abundance}
\label{psgabundace}
We found no particles with extreme O or C isotopic anomalies clearly indicative of a circumstellar and presolar origin. While there are several small particles among the 63 measured in the bulb of track C2052,74 that deviate more than 50\permil\xspace from the terrestrial O isotope ratios (Figure \ref{StardustObulb}), the large majority of presolar grains have more-anomalous O isotopic composition: only $\sim$7\% of presolar oxides and silicates in the presolar grain database \citep{hynes2009presolar} were measured to be $-200\permil<$\xspace\delox$<+200\permil$ and $-200\permil<$\xspace\deloy$<+200\permil$. Here, we define presolar grains as $|$\xspace\delox$|>$200\permil\xspace or $|$\xspace\deloy$|>$200\permil. We identified 0 presolar grains in our measurements. If we assume the simplest case that the number of presolar grains in a sample of grains from comet Wild 2 are Poisson-distributed with mean $\lambda$, then the Poisson single-sided lower limit for the estimate of $\lambda$ is given by:
% We found no O-anomalous presolar grains in the bulb section we measured from track C2052,74. If we assume the simplest case that presolar grains are Poisson-distributed with mean $\lambda$, then the Poisson single-sided lower limit for the estimate of $\lambda$ is given by:
\begin{equation}
\lambda \ge \frac{1}{2} \chi^2(\alpha;2k)
\end{equation}
and similarly, the Poisson single-sided upper limit for the estimate of $\lambda$ is given by:
\begin{equation}
\lambda \le \frac{1}{2} \chi^2(1-\alpha;2k+2)
\end{equation}
where $\chi^2$ is the inverse of the chi-square cumulative distribution function (we used \texttt{chi2inv} in Matlab 8.0.0.783) for confidence level 1-$\alpha$ and observed number of events $k$. We found 0 presolar grains, so setting $k=0$, we calculate the single-sided upper limit at the 95\% confidence level ($\alpha$=0.05) to be 3.0 presolar grains, which equates to an abundance of 4.8\% (3.0/63). (That is, the 95\% confidence upper bound of the presolar grain abundance in Wild 2 fines, based on our measurements, is 4.8\%). One presolar grain, a silicon carbide, has been found in the Stardust aerogel \citep{Brownlee:2009p6351}. Clearly, many more measurements need to be made to ascertain a meaningful statistical upper bound of presolar grains in the comet Wild 2 aerogel fines. 

The presolar grain abundance has been estimated from the Stardust foils, though care must be taken to account for presolar grain destruction caused by severe impact heating. \citet{Floss:2013p6164} estimated the presolar grain abundance in Stardust foil samples using high-speed analog shots of the Acfer 094 meteorite to account for presolar grain destruction. They found that the Acfer 094 impact residue was depleted in presolar grains compared to its nominal abundance and concluded that the same loss of presolar grains affected the Stardust foil samples. \citet{Floss:2013p6164} calculated the silicate+oxide presolar grain abundance in comet Wild 2 to be 600--830 ppm. This confidence interval represents only the range of Acfer 094 presolar grain abundance measurements, it does not take into account any statistical uncertainty. An estimate of the statistical uncertainty of these measurements, again assuming a Poisson distribution, yields a 95\% one-sided Poisson upper bound of $\sim$0.3\%, a much stricter upper bound than our measurements in the aerogel fines. \citet{Leitner:2010p6156} analyzed only small craters on the Stardust cometary foils (where dilution effects are small) and found one presolar grain, corresponding to a 2$\sigma$ (Poisson single-sided) upper limit of 0.8\% on the presolar grain abundance.

At this point, extremely high abundances of isotopically anomalous presolar grains in comet Wild 2 cannot be ruled out. A few thousand ppm is possible in the bulk rocky material from the comet, and the fines can theoretically be much more enriched in presolar grains (percent-level) if the comet is a mixture of primitive fine-grained material and processed coarse-grained material.

\subsection{Fine-grained material}
The range of O compositions seen here in the bulb of Track 74 span the range of all known Solar-System materials, from very $^{16}$O-rich (mostly in Mg-rich, Fe-poor particles) to very $^{16}$O-poor (in a high Fe region). Map \#3 contained particles which showed the largest variations in O isotopic composition. The $^{16}$O-poor, Fe-rich region in Map \#3 has O isotopic composition similar to cosmic symplectite \citep[\deloxy = $+180\permil$,][]{seto2008mineralogical}. However, cosmic symplectite is made of magnetite and pentlandite and the $^{16}$O-poor Wild 2 material is amorphous (Section \ref{remeasureaerogel}). This Fe-rich material could have been an Fe-bearing crystalline object, such as magnetite, in the comet before being amorphized by high-speed capture or the ion probe measurements prior to TEM analysis. It is thought that Acfer 094 cosmic symplectite formed by the oxidation and sulfurization of Fe metal grains through interaction with outer nebular water \citep{seto2008mineralogical}. The anomalously \ce{^{17,18}O}-rich Wild 2 material could have formed in a similar way before being incorporated into the comet.

%Map \#3 also contains small Mg-rich grains that are very \ce{^{16}O}-rich. These grains might be related to \ce{^{16}O}-rich LIME (low-iron, manganese-enriched) olivines as measured in Wild 2 grains by \citet{nakashima2012oxygen} or AOAs that are almost exclusively \ce{^{16}O}-rich \citep{krot2004amoeboid}. An Fe-rich region in Map \#3 was very \ce{^{16}O}-poor. This region was extracted by FIB, analyzed by TEM, and the Fe-rich grains were measured again for O isotopes by SIMS (Section \ref{remeasureaerogel}).

\begin{figure}[!ht]
\begin{center}
\includegraphics[width=\textwidth]{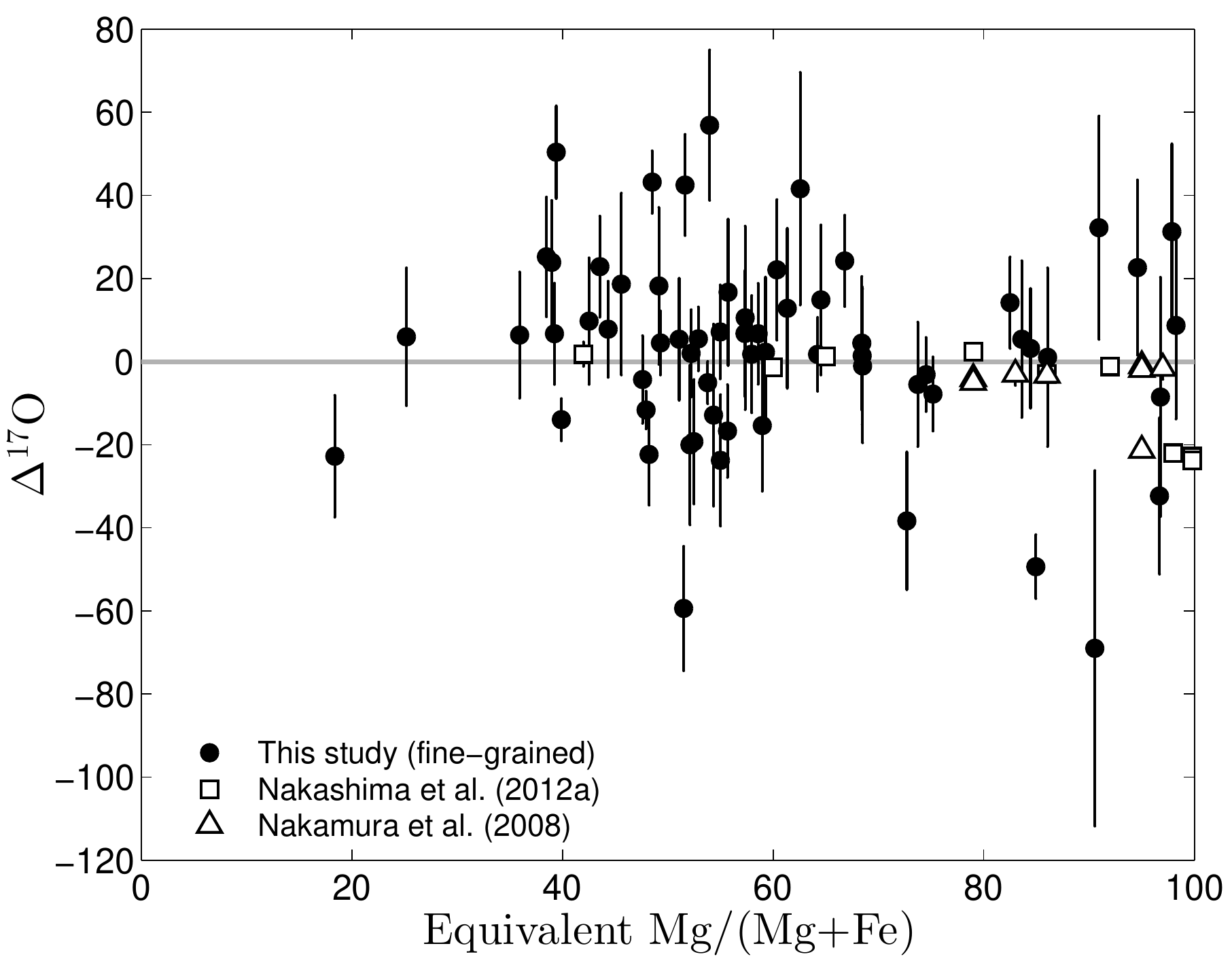}
\end{center}
\vspace*{-0.25in}
\caption{\Deloy vs. equivalent Mg/(Mg+Fe) (or, Mg\#) of 63 particles from the bulb of Stardust Track C2052,74 (filled circles: this study) compared with literature data (open squares: \citet{nakashima2012oxygen}, open triangles: \citet{Nakamura:2008p2349}). Error bars are 2$\sigma$. \label{fonumber}}
\end{figure}

\begin{figure}[!ht]
\begin{center}
\includegraphics[width=\textwidth]{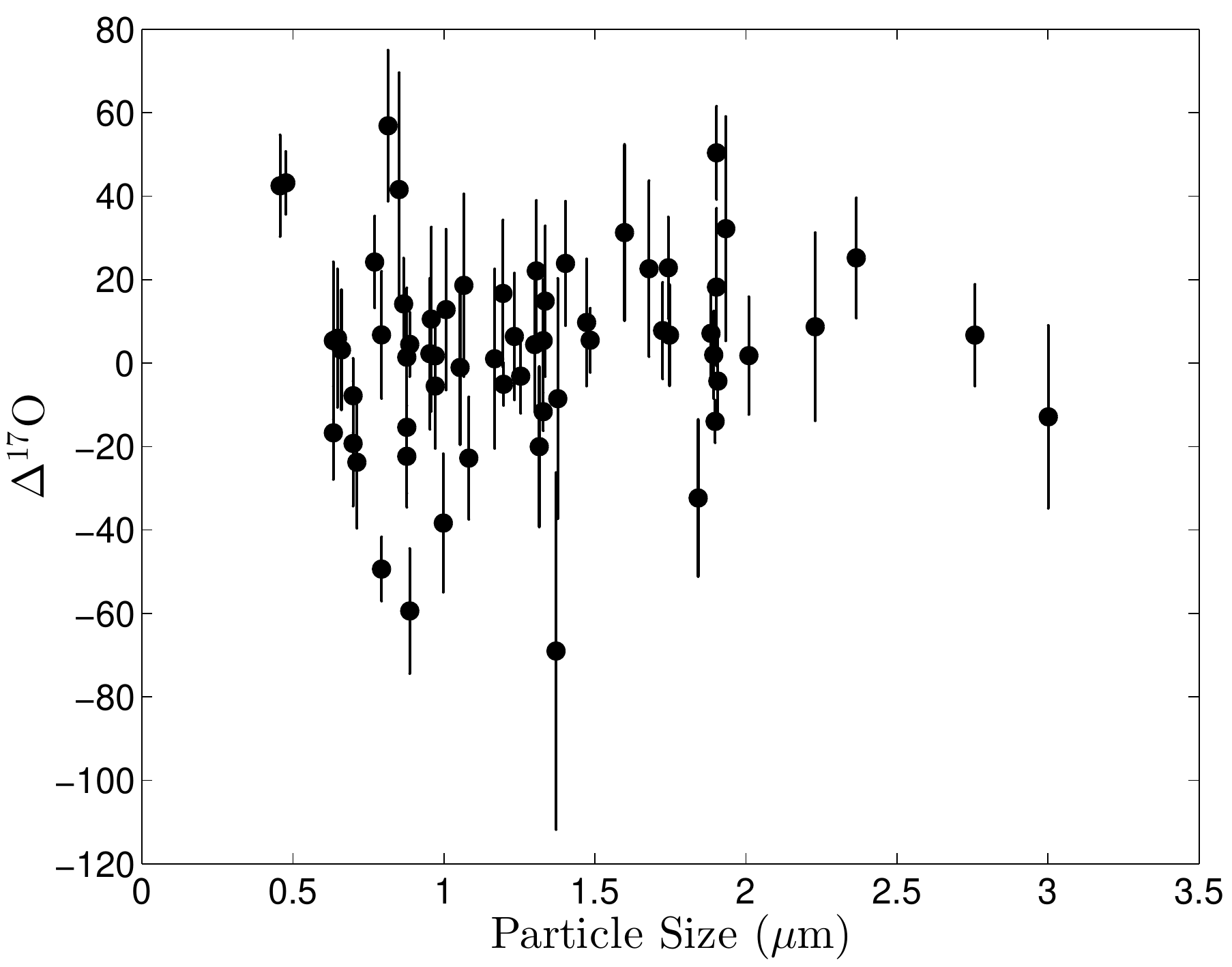}
\end{center}
\vspace*{-0.25in}
\caption{\Deloy vs. particle size of 63 particles from the bulb of Stardust Track C2052,74. Error bars are 2$\sigma$. \label{D17O_vs_particlesize}}
\end{figure}

Larger crystalline ferromagnesian silicates from comet Wild 2 show increasing \Deloy with increasing Mg/(Mg+Fe) \citep{nakashima2012oxygen}, similar to CR chondrite chondrules. Track C2052,74 fines have highly variable \Deloy which does not correlate well with Mg/(Mg+Fe) (correlation coefficient = $-0.14$) as shown in Figure \ref{fonumber}, or with particle size (correlation coefficient = $0.07$) as shown in Figure \ref{D17O_vs_particlesize}. The error- and size-weighted bulk \Deloy of the Track C2052,74 fines is 4.5$\permil \pm 2.9\permil$ (2$\sigma$).

The fine-grained material in this track appears to sample a much broader range of O composition than seen in larger terminal particles. The standard deviation of \delox of the 63 particles in our Stardust bulb measurements is 49$\permil$, compared to the $\sim$7$\permil$ standard deviation of our coarse-grained measurements (albeit from a much smaller sample size). From our isotopic measurements and mineralogical studies of bulb fines by \citet{Stodolna:2012p5651}, we conclude that the fine-grained Wild 2 material is not simply a smaller-size representation of the larger Wild 2 grains.

\citet{nakashima2012high} measured the O isotopic composition of an anhydrous interplanetary dust particle (Z17) that showed distinct \Deloy between its fine-grained ($-4.3 \pm 0.9$, 2$\sigma$) and coarse-grained ($-0.4 \pm 1.4$) components. The fine component of anhydrous interplanetary dust may have sampled different O isotopic reservoirs than the larger mineral grains, analogous to our measurements of the fine- and coarse-grained components of comet Wild 2. 

To understand the provenance of the fine-grained material from comet Wild 2, it is also useful to look at our measurements in the context of asteroidal fines. If fine-grained material in the early Solar System was ubiquitous at all radii, we would expect the Wild 2 fines to be similar to the matrix of primitive meteorites. If, instead, the Wild 2 fines sampled a different reservoir of O than asteroids did, we would expect to see differences in their O composition. In particular, is the \textit{diversity} of O compositions sampled by the Wild 2 fines seen in the fines in another cometary or asteroidal parent body? Interplanetary dust particles are samples of a large number of parent bodies, not a single asteroid or comet, so in the following discussion we will compare our Wild 2 measurements with fine-grained material in meteorites.

%other analyses of  it to previous measurements of fine-grained material measured in meteorites. We compare our O isotope measurements of the cometary material with matrix in Vigarano,  a relatively pristine carbonaceous chondrite, and alumina grains separated from unequilibrated ordinary chondrites.

\citet{Kunihiro:2005p6320} measured the O isotopic composition of the Vigarano meteorite, a CV3 chondrite that has undergone mild thermal and aqueous alteration \citep{Lee:1996p6353}. They determined that the upper bound standard deviation (limited by the measurement's statistical error) of the very fine-grained ($<$500 nm grains) groundmass matrix of Vigarano is 5$\permil$ in \delox. This is much smaller than the 49$\permil$ standard deviation from our measurements, indicating that the fine-grained material in comet Wild 2 likely sampled a more diverse source of O isotopes than the fine-grained material in the CV parent asteroid. However, the fines in CV3 meteorites likely experienced weak thermal and aqueous processing on the asteroid which may have partially equilibrated O isotopes and produced a relatively narrow distribution of O isotopic compositions. There was likely no significant thermal or aqueous processing on comet Wild 2 which would produce a narrow, Vigarano-like distribution of O isotopes in the Wild 2 fines. A study of the O isotope composition of individual matrix grains of primitive meteorites that experienced minimal alteration, such as the CR chondrites Queen Alexandra Range (QUE) 99177 and Meteorite Hills (MET) 00426 \citep{abreu2010early} or the ungrouped carbonaceous chondrite Acfer 094 \citep{greshake1997primitive}, would make for a more appropriate comparison with the comet Wild 2 fines.

Presolar grains and small refractory matrix grains in primitive meteorites escaped homogenization during chondrule formation. \citet{takigawa2014morphology} measured the O isotopic composition of 107 alumina grains ($\sim$95\% were corundum) 1 $\mu$m and smaller from unequilibrated ordinary chondrites. They found a very large spread in O composition for the solar corundum grains (Figure 9 of \citet{takigawa2014morphology}), including very $^{16}$O-rich and $^{16}$O-poor grains. The standard deviation of \delox for these grains is $\sim$32$\permil$. These measurements of corundum grains in UOCs are similar to the spread of O compositions we measured in the fine-grained component of comet Wild 2. 

% This is probably a biased (high) estimate of the presolar grain abundance, so leave it off:
%Nine of 107 corundum grains showed O isotopic anomalies consistent with a presolar origin. The 95\% Poisson single-sided lower-limit (see Section \ref{psgabundace}) on nine grains is 4.7 grains, which gives a 95\% confidence lower-limit abundance of 4.4\%. The single-sided 95\% upper limit for the presolar grain abundance in our measurements of the comet Wild 2 fine-grained material is 4.8\%  (Section \ref{psgabundace}). Therefore it is likely that the corundum grains measured by \citet{takigawa2014morphology} have a higher presolar grain abundance than the Wild 2 fines, however, the estimates are currently not distinguishable with high confidence due to limited statistics. 

\citet{makide2009oxygen} measured 109 larger (1--5 $\mu$m) corundum grains and found them to have $^{16}$O-rich compositions (mean \Deloy = $-22\permil$) with 15$\permil$ standard deviation in \delox. These corundum grains appear to be more homogenous and $^{16}$O-rich than the smaller corundum grains measured by \citet{takigawa2014morphology}. At larger size scales, refractory inclusions like AOAs and CAIs also have relatively uniform $^{16}$O-rich compositions. The larger objects, CAIs, AOAs, and larger corundum grains, likely formed in the solar nebula close to the young Sun. (Some corundum grains from carbonaceous chondrites formed via hydrothermal alteration on a parent body and have O isotopic compositions consistent with the terrestrial fractionation line  \citep[e.g.][]{maruyama2008hercynite}). The small corundum grains measured by \citet{takigawa2014morphology} may be circumstellar grains, but lack the nucleosynthetic isotope anomalies that are typically used to label them as presolar. (The isotopically anomalous grains measured by \citet{takigawa2014morphology} mostly had rough surface morphology possibly caused by energetic particle irradiation in interstellar space. The corundum grains with rough surfaces but without extreme isotope anomalies may have had similar origins and histories as the anomalous grains.) These corundum grains could be a component of the solids inherited from our parent molecular cloud, reflecting the diverse O isotopic compositions of the primitive constituents of the Solar System. Corundum grains survived processing on the UOC parent body and were able to retain their original, diverse O isotopic composition because self-diffusion of O in corundum is very slow \citep{heuer2008oxygen}. Other small, less robust molecular cloud grains were destroyed or their O isotopic signatures were erased by nebular and/or asteroidal processes and are not found in meteorites. 

% for solids that were formed and subsequently processed in the Solar System

Our measurements of a very wide range of O isotopic compositions in the Wild 2 fine-grained material, covering the entire range of known Solar-System materials, can be explained by one or a combination of the following two scenarios: 1) The Wild 2 fines sampled many oxygen reservoirs from the inner solar nebula, or 2) the fines are representative of the raw starting materials, inherited from the parent molecular cloud, from which the inner Solar System material formed.

\subsubsection{Scenario \#1: Comet Wild 2 fines sampled many oxygen reservoirs from the inner Solar System}
In this scenario, the comet Wild 2 fines formed in the inner solar nebula and were transported to the outer nebula where they were incorporated into the comet. The fine-grained material in the outer Solar System at the time of the comet's formation was mostly of foreign origin, implying that the \textit{transport of inner Solar System material to the outer solar nebula was efficient}. Other fine-grained material that was present in the time and place where the comet formed, i.e. surviving molecular cloud solids, would have been incorporated into the comet as well. But if most of the comet Wild 2 fines are of foreign provenance, \textit{the amount of surviving molecular cloud material would be small compared to inner Solar System material}. 

We can make testable predictions if the Wild 2 fines are from the inner Solar System. Presolar grains are more likely to survive in the cold outer regions of the solar nebula compared to the inner nebula where they are destroyed by nebular processing. Therefore, the presolar grain abundances of inner Solar System material should be less than unprocessed outer nebula material. In this scenario, \textit{the presolar grain abundances in Wild 2 fines should be comparable to the abundances in primitive meteorites}, which incorporated inner Solar System material but did not destroy presolar grains through heating or aqueous alteration on the parent asteroid. However, as described in Section \ref{psgabundace}, the statistical uncertainty of the presolar grain abundance in a given parent body can be very large due to limited sampling. 

Volatile species are more easily retained in the cold outer nebula compared in the inner nebula. Most meteorites are depleted in volatile elements relative to the Sun, except for CI chondrites which escaped high-temperature processing and have chemical composition similar to the solar photosphere. The volatile depletion patterns seen in the various chondrite classes were likely caused by variable heating of material in the protoplanetary accretion disk \citep{davis2006volatile}. If Wild 2 fines were sourced from the inner Solar System, they would also be \textit{depleted in volatiles compared to CI chondrites}.

\subsubsection{Scenario \#2: Comet Wild 2 fines are representative of the raw starting materials, inherited from the parent molecular cloud, from which the inner Solar System material formed}
Conversely, if the fines in Wild 2 were not transported from the inner Solar System, we would draw different conclusions about the nature of material in the solar nebula at large distances from the Sun, and make different predictions about the presolar grain abundances and volatile depletion in Wild 2 fines. In this scenario, \textit{the fine component would be dominated by grains inherited from the Solar System's parent molecular cloud} (e.g. circumstellar grains with nucleosynthetic signatures, amorphous interstellar silicates).  \textit{Abudances of isotopically anomalous presolar grains would be high compared to primitive meteorites} and \textit{volatiles would not be significantly depleted compared to CI chondrites}.

%(``The Wild 2 fines sampled the all of the oxygen reservoirs of inner Solar System materials''), the following would also have to be true:
%\begin{enumerate}
%\item The transport of all types of inner Solar System to the scattered disk (where comet Wild 2 formed) was extremely efficient.
%\item The amount of surviving molecular cloud material is small compared to the foreign inner Solar System material.
%\item Presolar grain abundances of SiC and oxides in the Wild 2 fines are small compared to these abundances CI chondrites (which escaped high-temperature processing).
%\item The bulk chemical composition of Wild 2 fines are depleted in volatiles compared to CI chondrites. 
%\end{enumerate}

% For Scenario \#2 to be true, the following would also have to be true:
% \begin{itemize}
% \item The fine component of material in the scattered disk is dominated by material inherited from the parent molecular cloud.
% \item Presolar grain abundances of SiC and oxides are high, comparable to these abundances in CI chondrites.
% \item The bulk chemical composition of Wild 2 fines is similar to CI.
% \end{itemize}

Our current measurements do not strongly support Scenario \#1 or \#2. Presolar grain abundances and the bulk chemical composition of the Wild 2 fines are currently too imprecise to compare to measurements of primitive meteorites.

The diverse O isotopic compositions of the fines were not equilibrated on the comet because it was kept in cold storage beyond the orbit of Neptune since the Solar System's formation. Our measurements of diverse O isotopic compositions agree with TEM studies of Wild 2 fines by \citet{Stodolna:2012p5651} which found unequilibrated, heterogenous particles with approximately chondritic average chemical composition.

The larger terminal particles collected by Stardust have high-temperature mineralogy and $^{16}$O-poor compositions that indicate they formed in the inner Solar System and were transported to the scattered disk where they were incorporated into comet Wild 2. The lack of $^{26}$Al in four Wild 2 fragments measured so far \citep{matzel2010constraints,ishii2010lack,Ogliore:2012p5583,nakashima2014lack} tells us that this transport happened relatively late, more than 3 Myr after the onset of CAI formation. These coarse-grained inner Solar System objects combined with primitive dust and ices in the cold, sparse, outer Solar System, and eventually accreted together to form comet Wild 2.

It should be noted that our measurements of the 63 bulb particles from comet Wild 2 represent only a small section of the bulb of only one of hundreds of tracks returned by the Stardust mission. Similarly, there have only been $\sim$25 larger Wild 2 particles measured for O isotopes \citep{McKeegan:2006p936,Nakamura:2008p2349,2008M&PSA..43.5308M,nakamura2011nanometer,nakashima2012oxygen}. The diversity seen in these cometary samples so far underscores the need for more O isotope and chemical analyses of both the terminal particles and the fines. There have been many more analyses of a given individual type of chondrite, which represents a much more homogeneous parent body than comet Wild 2. Many more tracks need to be analyzed before we can confidently understand the composition of this comet.

% little variation (mean \Deloy = $-23\pm$7, 2$\sigma$).

% fewer presolar grains (1 of 103, single-sided 95\% upper limit = 4.4\%)

% Corundum is highly refractory and is one of the first minerals to condense out of a gas of solar composition \citep[e.g.][]{krot2012heterogeneous}.

%Compare to Takigawa's O isotopes of corundum grains.

%Compare to QUE99177 MET00426 Acfer 094?? Is the data out there?

\section{Conclusions}
\label{conclusions}
We measured the O isotopic composition of seven $>$2~$\mu$m fragments extracted from five Stardust aerogel tracks, and 63 small particles ($<$2~$\mu$m) measured in compressed aerogel from the bulb of track C2052,74. The larger fragments have O isotopic compositions mostly consistent with the terrestrial fractionation line. The fine-grained material shows very diverse O isotopic compositions that span the range of all Solar System materials, and are similar to the O isotope compositions seen in $<$1~$\mu$m corundum grains extracted from unequilibrated ordinary chondrites. We conclude that the larger particles were formed in the inner Solar System from an evolved O isotopic reservoir, and transported to the outer Solar System. These larger particles then combined with unequilibrated dust of diverse O isotopic composition, inherited from the Solar System's parent molecular cloud or from a diverse sample of inner Solar System oxygen reservoirs, to form the rocky component of comet Wild 2.

\section{Acknowledgements}
The authors thank Daisuke Nakashima, Larry Nittler, and an anonymous reviewer for their suggestions that helped improved this paper. The National Center for Electron Microscopy is supported by the Director, Office of Science, Office of Basic Energy Sciences, of the U.S. Department of Energy, under Contract No. DE-AC02-05CH11231.
This work was supported by NASA grants NNX07AM62G (G.R.H.), NNX07AM67G (A.J.W.), and NNX14AF24G (R.C.O.). 

%\begin{itemize}
%\item Compare (plot?) fine-grained material O composition with Kunihiro et al ``Microscopic oxygen isotopic homogeneity/heterogeneity in the matrix of the Vigarano CV3 chondrite'' matrix grains (` `The groundmass is isotopically homogeneous and 16O-poor (Fig. 3c). The standard deviation of d18O SMOW in the ground- mass area is estimated from binned pixels and calculated to be 5 $\permil$. T '') and Takigawa et al and ``Morphology and crystal structures of solar and presolar Al2O3 in unequilibrated ordinary chondrites'' alumina grains.
%\end{itemize}

\clearpage
%\section{References}
\vspace{-2pc}
\bibliographystyle{model2-names}
\bibliography{references}
\end{document}